\documentclass[11pt]{article}

\ifdefined\pdfminorversion \pdfminorversion=7 \fi

\usepackage[margin=1in]{geometry}
\usepackage{amsmath, amssymb, amsfonts}
\usepackage{graphicx}
\usepackage{float}
\usepackage{booktabs}
\usepackage{siunitx}
\usepackage[final,protrusion=true,expansion=false]{microtype}
\usepackage{xcolor}
\usepackage[numbers,sort&compress]{natbib}
\usepackage[colorlinks=true, linkcolor=blue, citecolor=blue, urlcolor=blue]{hyperref}
\usepackage{url}
\usepackage{tikz}
\usetikzlibrary{arrows.meta, positioning, calc, fit}
\newcommand{\VpiL}{V_\pi L}
\newcommand{\Cj}{C_j}
\newcommand{\neff}{n_\mathrm{eff}}
\newcommand{\Tidy}{Tidy3D}
\newcommand{\PhotonForge}{PhotonForge}

\title{Autonomous agentic design for photonics}
\author{%
  Prashanta Kharel \and
  Amin Khavasi \and
  Xinzhong Chen \and
  Tyler W. Hughes\\
  \small Flexcompute Inc.\\
  \small \texttt{\{prash, amin.khavasi, tom, tyler\}@flexcompute.com}%
}
\date{\today}

\begin{document}
\maketitle

\begin{abstract}
\noindent
We introduce an automated, agent-driven approach to the design of photonic devices.
We instruct large language models (LLMs) to solve photonic design problems, given access to software tools for performance evaluation (through numerical simulations) and quantitative acceptance criteria (e.g., fabrication rules, geometric constraints, physical-consistency checks).
Within this context, agents run autonomous design loops (propose, simulate, evaluate, iterate) and generate devices with state-of-the-art performance.
We demonstrate this approach in two stages:
First, we run it individually on four canonical problem classes in photonic chip design:
a) passive components (waveguide bends, splitters, crossings, etc.);
b) active devices (silicon microring modulators (MRMs));
c) radio-frequency (RF) devices (traveling-wave electrodes for a Mach-Zehnder modulator (MZM));
d) chip layout (electrical routing).
Then, we combine the previous studies in one demonstration to produce a silicon photonic modulator, incorporating layout, charge transport, optical mode, and RF electrode design.
The approach generalizes to any problem that combines a numerical simulator with performance criteria that an LLM can evaluate.
\end{abstract}

\section{Introduction}
\label{sec:intro}

The field of photonics uses light to deliver signals with high bandwidth, low propagation
loss, and high propagation speed for various technologies.
Applications include optical interconnects for communications
\citep{reed2010silicon, bogaerts2018silicon, torrijos2026datacenter},
LiDAR \citep{li2022lidar}, AR/VR display optics
\citep{xiong2021arvr}, and optical computing
\citep{margalit2021perspective, kazanskiy2022optical}. With manufacturing
techniques borrowed from the semiconductor industry, these capabilities are often
realized on-chip in photonic integrated circuits (PICs), enabling compact,
low-cost, and mass-manufacturable devices.

However, designing photonic devices is a slow and manual process. The conventional
process involves studying the existing literature, setting up numerical simulations,
optimizing parameters, interpreting results, and iterating, with a human engineer guiding each
step. Within a single device, different components often
require different physics: passive devices such as waveguides rely on electromagnetic
simulation (optical mode solvers, finite-difference time-domain [FDTD] simulation, eigenmode expansion
method [EME]), active components such as modulators involve semiconductor
carrier transport simulation, and RF electrodes require microwave field analysis. Combining these
components into a circuit adds further requirements: circuit-level simulation,
physical routing, and system-level co-design \citep{bogaerts2018silicon}.
Furthermore, the fabrication process after simulation introduces variation often not
accounted for in the simulation models, requiring multiple rounds of tape-outs to verify and correct initial device designs.

As a result, individual components alone can take weeks or months to design and entire devices even longer, at enormous cost.
GPU-accelerated simulation can speed up the evaluation part of this process \citep{hughes2021pathway},
but often thousands or more simulations are needed to complete a design, and results need to be coordinated across different roles in an engineering team, including photonic device designers, layout engineers, system engineers, and measurement engineers.
Some aspects of the component design process can be automated through techniques like gradient-based ``inverse design''
\citep{molesky2018inverse, hughes2018adjoint, shi2026adjoint} optimization, but these cannot handle discrete
decisions such as changes in design parameterization or architecture, cannot form
hypotheses about failure mechanisms, and still require a human in the loop to
coordinate across physics and tools.

AI agents based on large language models (LLMs) are natural candidates
to translate natural language into technical specifications, but also
increasingly to drive simulation and design loops autonomously.
Recent work has begun to apply them to photonic design. Metasurface and metamaterial studies have shown that agents
can propose, evaluate, and revise nanophotonic geometries under simulator or
surrogate feedback \citep{lupoiu2025metachat, huang2026selfevolving,
wu2026metadesigner, lu2025agentic, kim2025nanophotonic, mahlau2025marl}. In
PIC design automation, PhIDO \citep{sharma2025aiagents} demonstrated
natural-language-driven generation of PDK-based chip layouts with layout and
circuit checks, and a complementary benchmark for evaluating LLMs on PIC
design tasks has been proposed \citep{wu2025picbench}. Classical
electronic-photonic design-automation work such as PoLaRIS
\citep{zhou2025polaris} addresses device inverse design and physical layout
generation without an LLM agent. These efforts make important contributions to automating the design process,
but do not yet produce true end-to-end agent-driven design across multiple photonic device classes,
combinatorial chip-level routing, and a coupled multiphysics workflow.
A detailed comparison against these and other representative efforts, across five
axes for agentic PIC design, is given in Table~\ref{tab:positioning}
(\S\ref{si:positioning}).

In this work, we demonstrate fully automated, agent-driven photonic design in two stages.
First, we apply the approach to four critical problem classes in photonic chip design: passive components (electromagnetic simulation of waveguides, splitters, bends, etc.), active devices (drift-diffusion simulation of a PN-junction phase shifter under bias), RF electrodes (3-D electromagnetic simulation of a traveling-wave electrode), and chip-level routing.
In each case, we define success criteria as functions of the evaluator's outputs (simulation results, or design-rule check counts for routing) and any fabrication or geometric constraints, and the agent iterates autonomously until the criteria are met.
Second, we combine all of the previous elements in one outer loop to design a silicon Mach--Zehnder modulator (MZM) \citep{reed2010silicon, khavasi2026cps}, a widely-used silicon photonic device for high-speed electro-optic modulation.
In this demonstration, layout, charge transport, optical mode, and RF electrode are coupled, and the agent co-optimizes them in one loop, reproducing and extending a peer-reviewed device.
The success of such an approach suggests high potential value in other engineering fields where agents can drive numerical simulation tools and engineers can define evaluation criteria.

\section{The agentic photonic design loop}
\label{sec:workflow}

Autonomous agents that research and work toward a goal were recently
popularized by self-improving machine-learning systems
\citep{karpathy2026autoresearch} and have since been applied across the
sciences, from computational chemistry \citep{boiko2023autonomous} to
autonomous materials synthesis \citep{szymanski2023autonomous} and
mathematical discovery \citep{romeraparedes2024funsearch}.
In this paradigm, an agent generates, evaluates, and revises candidates against measurable validation criteria using software tools.
We apply the same idea to photonic chip design, where the evaluator is a
numerical simulator with verifiable outputs and we include constraints related to the PIC manufacturing process.
The agent itself is a general-purpose commercial coding assistant, such as Claude from Anthropic or Codex from
OpenAI, so the approach uses widely available tools rather than a bespoke model.

\begin{figure}[H]
\centering
\begin{minipage}[c]{0.42\textwidth}\centering
\includegraphics[width=\linewidth]{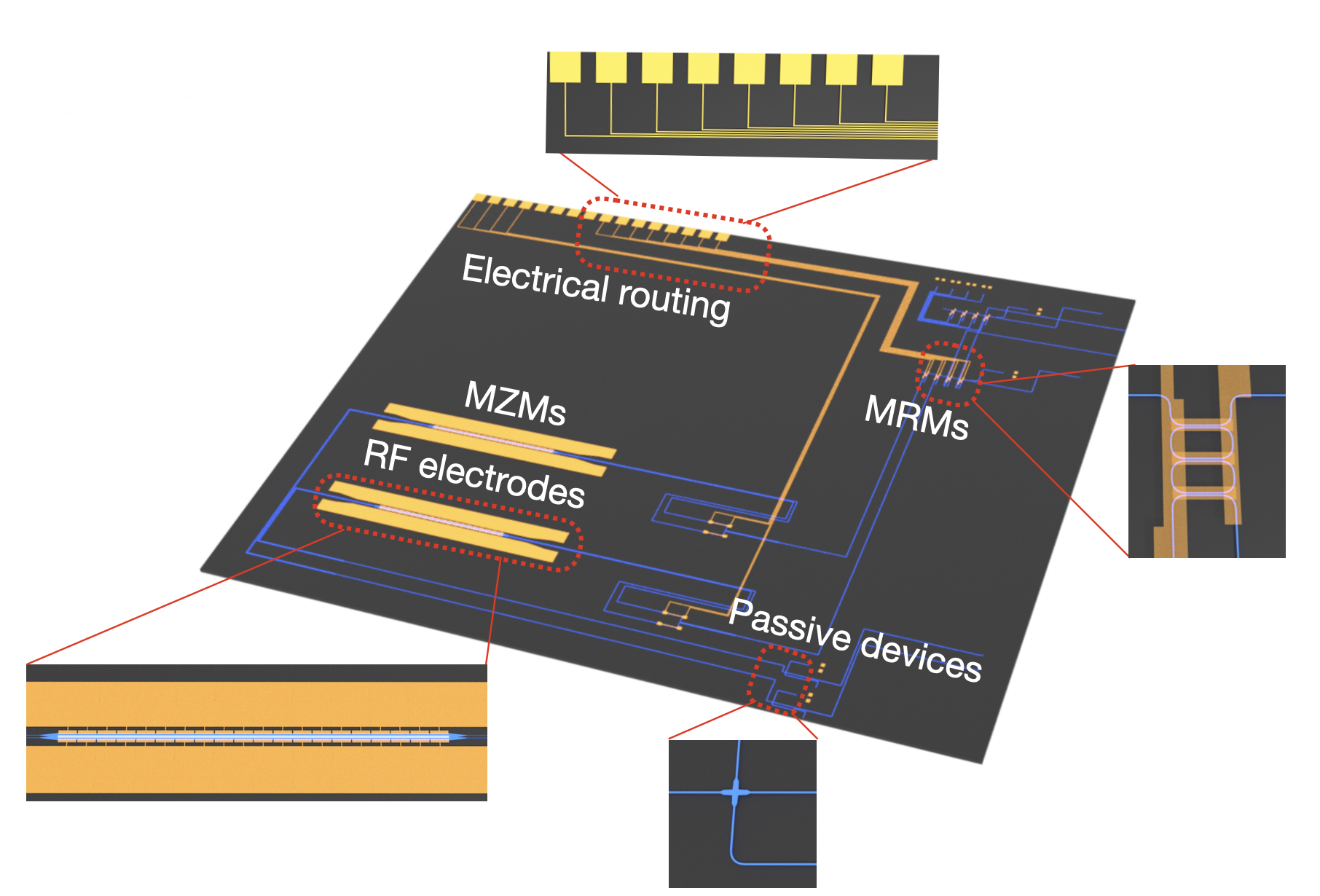}\\[3pt]
\textbf{(a)}
\end{minipage}\hfill
\begin{minipage}[c]{0.525\textwidth}\centering
\includegraphics[width=\linewidth]{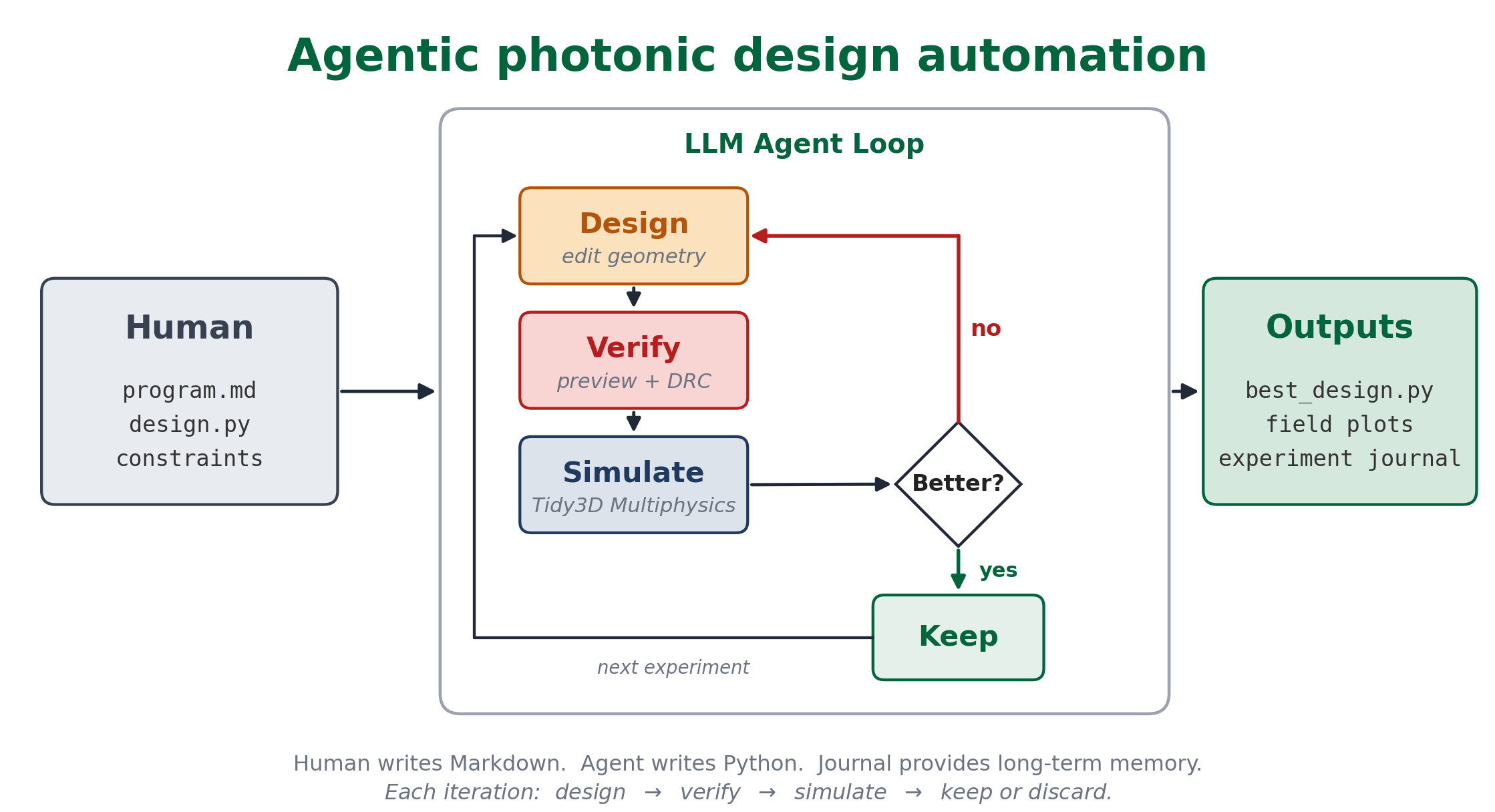}\\[3pt]
\textbf{(b)}
\end{minipage}
\caption{\textbf{Agentic photonic design automation.}
\textbf{(a)}~Photonic chip illustrating the design problems addressed
here, with zoom-in views of each: electrical routing, MRMs, passive
devices, MZMs, and RF electrodes. \textbf{(b)}~The agentic photonic
design automation loop, showing the design, verification, and simulation
cycle that the agent enters to iterate toward a design goal set by
humans under a set of constraints.}
\label{fig:design_loop}
\end{figure}

We start by giving the agent a photonic design problem, a figure of merit, a set of
design constraints, and the ability to run simulations.
Next, we let it iterate toward an acceptable design on its own.
The engineer provides an instruction document with the device target
and constraints, a simulation file the agent edits to set geometry, material
parameters, and routing, starting scripts for simulation, geometry preview,
and design-rule checking (DRC)~\citep{schubert2022inverse}, and a running journal that serves as the
agent's long-term memory. The agent then proposes a change, checks
constraints, runs the simulation, evaluates the figure of merit, and keeps or
reverts the change, repeating until the criteria are met
(Fig.~\ref{fig:design_loop}). The workflow presented
here,\footnote{Public reference implementation:
\url{https://github.com/flexcompute/autophotonicdesign}.} including
explicit steps for literature review and hypothesis generation, was
found empirically to work well for these problems
(\S\ref{si:passive}--\ref{si:multiphysics}). However, this approach
generalizes to any workflow, including simply specifying the performance and
acceptance criteria and allowing the agent to propose its own strategy to
meet the design objective.

The simulation produces numerical outputs that can be scored for performance and checked against well-defined pass/fail criteria,
which gives the agent a reliable signal without a human in the inner loop. We
use DRC broadly, to mean any constraint that can be encoded as a pass/fail
test, whether geometric, physical, or fabrication-related. These guardrails
keep the design exploration on physically meaningful candidates, and an
initial literature review gives the agent a strong starting point.

\section{Results}
\label{sec:results}

We now demonstrate the application of this agentic photonic design
automation loop across four photonic design problems spanning passive
and active device design, RF electrode design, and chip-level electrical
routing, and then to an end-to-end multiphysics simulation and design of
a coplanar stripline silicon photonics modulator. In each case, the
engineer sets up the initial design problem, and the agent runs the
loop autonomously. For brevity, we include the setup details and
supporting figures per-problem in the Supplementary Information.

\subsection{Passive components}
\label{sec:passive}
\label{sec:passive_others}

First, we apply the workflow to five passive device classes: a compact waveguide bend in
silicon nitride (SiN), and on a \SI{220}{\nano\metre} silicon-on-insulator
(SOI) platform a 1$\times$2 splitter, a waveguide taper, a focused grating
coupler, and a waveguide crossing. The agent reached excellent
performance on all five (\S\ref{si:passive}): bending loss of
\SI{0.109}{\decibel} at \SI{12}{\micro\metre} radius, splitter insertion loss
of \SI{0.015}{\decibel}, taper insertion loss of \SI{0.114}{\decibel} over
\SI{6}{\micro\metre}, grating-coupler peak coupling loss of
\SI{2.89}{\decibel}, and crossing insertion loss of \SI{0.11}{\decibel}.

\begin{figure}[H]
\centering
\includegraphics[width=\textwidth]{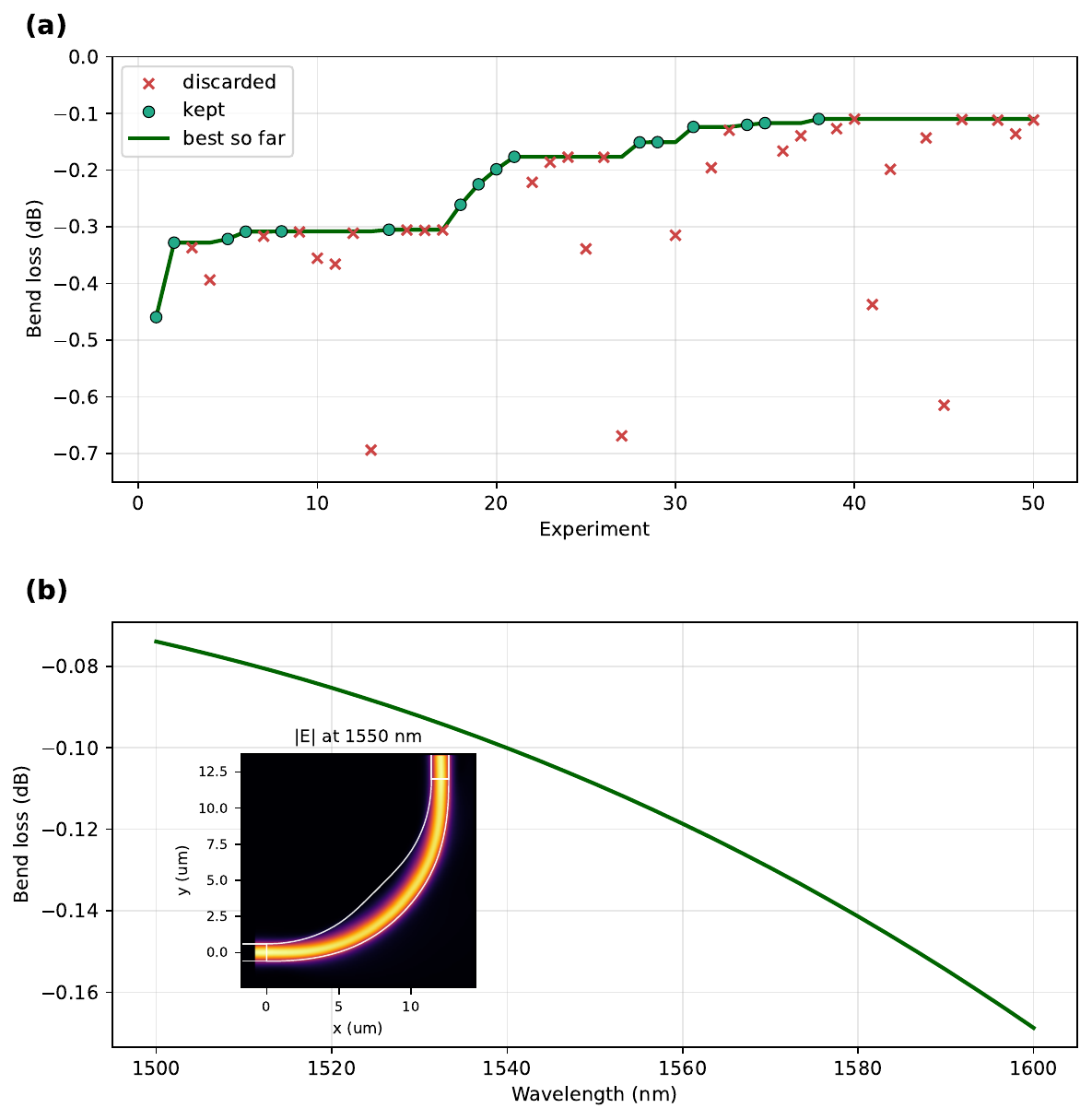}
\caption{\textbf{Passive bend.} \textbf{(a)}~Bend loss versus experiment over
the 50-experiment budget (red crosses: discarded attempts; green circles:
promoted to the running best; green curve: best-so-far envelope). 
\textbf{(b)}~Wavelength dependence of the optimized bend loss, with the
$|E|$ field profile at \SI{1550}{\nano\metre} (inset). White curve outlines the shape of the bend.}
\label{fig:bend}
\end{figure}

We focus here on the results from the waveguide bend as it is a ubiquitous component in PIC devices.
A tight bend radius is preferred for dense on-chip integration.
However, if the radius is too small, especially in a low-index-contrast platform such as SiN, the bend loss can be significant. Therefore, designing a compact low-loss bend is a challenging
and practically important problem. Given a basic circular bend baseline and a budget of 50 iterations, the agent following the design automation loop reached \SI{97.51}{\percent} fundamental-mode
transmission (\SI{0.109}{\decibel} bend loss) at a fixed \SI{12}{\micro\metre}
radius (Fig.~\ref{fig:bend}), a roughly $4\times$ reduction in bend loss relative to the
\SI{89.96}{\percent} (\SI{0.460}{\decibel}) circular baseline.

We observed that the agent modified the bend shape across the design iterations, first replacing the initial circular bend
with an Euler-circular-Euler bend and then adding a width taper
and an inward radial offset (quantitative details in \S\ref{sec:passive_results}). In between the major shape changes, the agent also performed small parameter sweeps to optimize the parameter values.
The final design achieves state-of-the-art loss level at this radius compared to the literature
\citep{cherchi2013dramatic, vogelbacher2019analysis, bahadori2019universal}
(\S\ref{si:passive}). The agent's design is deeply inspired by physical intuition rather than black-box optimization. The Euler-circular-Euler bend has a smoother curvature evolution than a simple circular bend. Width tapering and 
radial offset provides a better matching between the straight waveguide mode and the bend waveguide mode, ensuring a continuous mode profile transition throughout the bend. 
Brute-force optimization methods like particle swarm and gradient-based inverse design methods \citep{molesky2018inverse, hughes2018adjoint, shi2026adjoint} are not invoked here, but the agent could in principle invoke them autonomously if needed.

\subsection{Active devices}
\label{sec:active}

Next, we focus on the design of a silicon photonic modulator, which
encodes data from the electrical domain to optical signals. Silicon
photonic modulators utilizing a PN-junction phase shifter are the
workhorse of high-speed electro-optic links, and their design grows harder
as systems push beyond \SI{200}{Gb/s} per wavelength. In this regime, a low
modulation-efficiency product $\VpiL$ and a low junction capacitance $\Cj$
are needed as $\VpiL$ sets the drive voltage while $\Cj$ sets the
RC-limited bandwidth and the energy per bit. We asked the agent to minimize
$\VpiL \cdot \Cj$ for a silicon microring modulator (MRM) with a lateral PN phase
shifter, coupling a drift-diffusion charge simulation and a per-bias
optical-mode solve inside each evaluation, with no prior knowledge of the
achievable trade-off.

\begin{figure}[H]
\centering
\begin{minipage}{0.46\textwidth}\centering
\includegraphics[width=\linewidth]{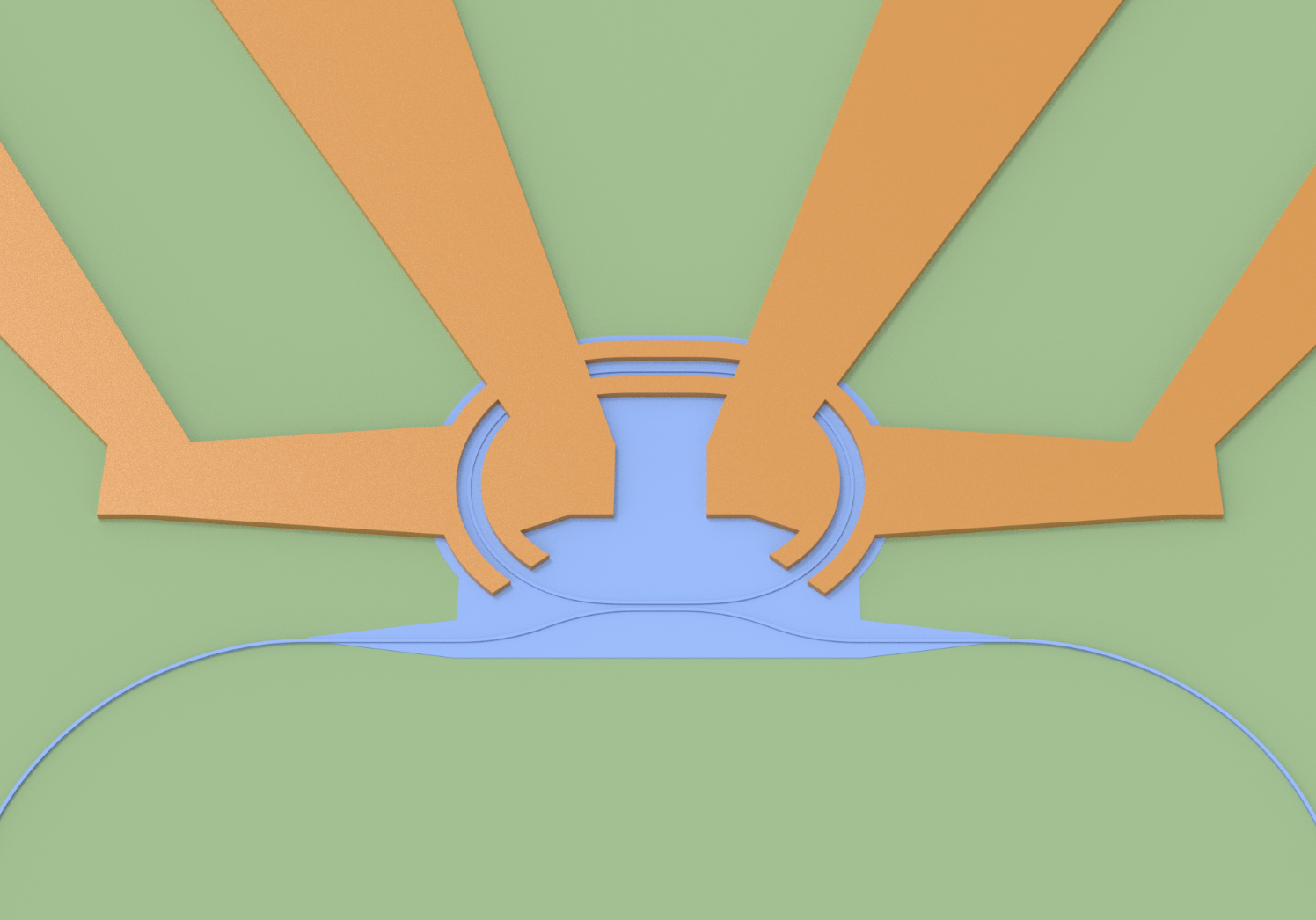}\\[2pt]\textbf{(a)}
\end{minipage}\hfill
\begin{minipage}{0.46\textwidth}\centering
\includegraphics[width=\linewidth]{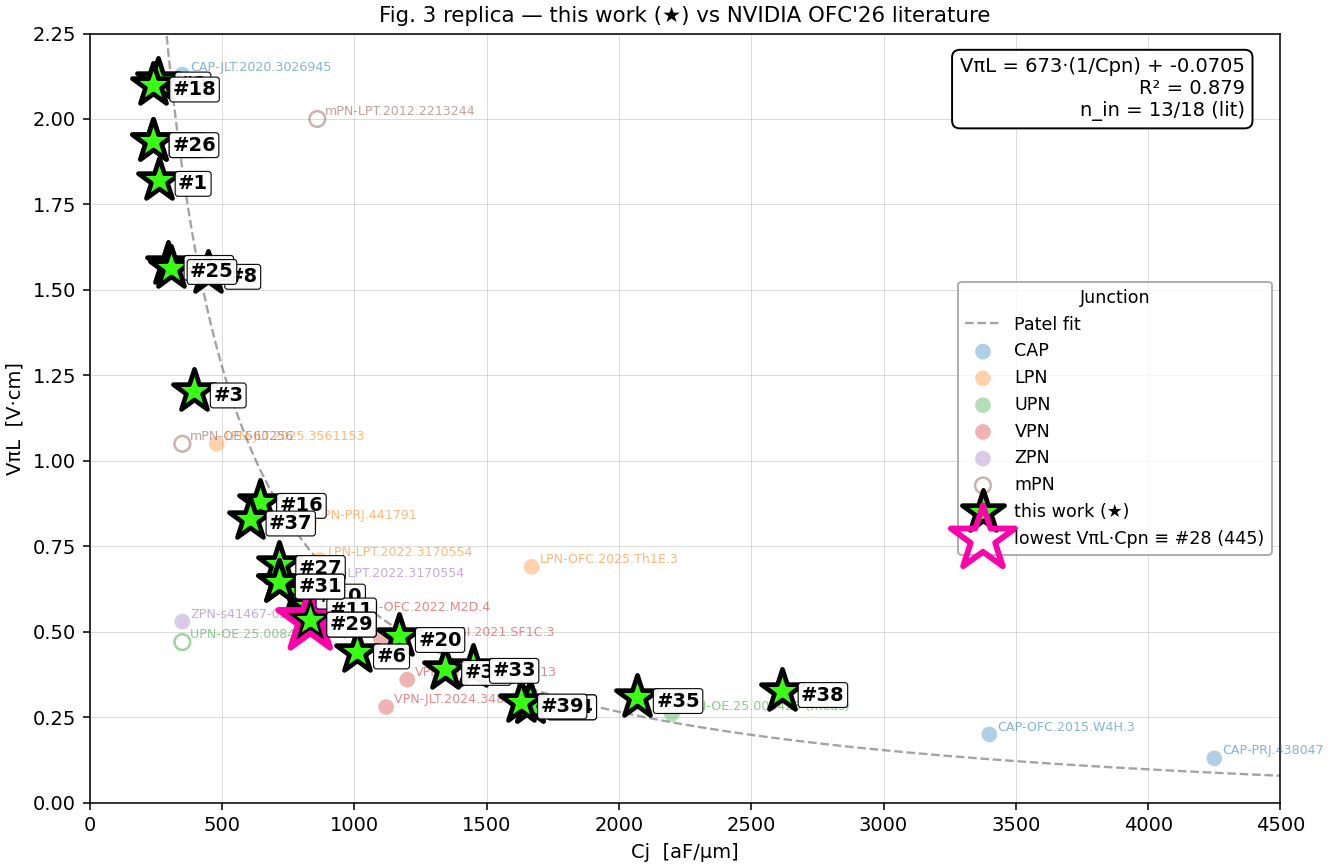}\\[2pt]\textbf{(b)}
\end{minipage}\\[0.6em]
\includegraphics[width=0.92\textwidth]{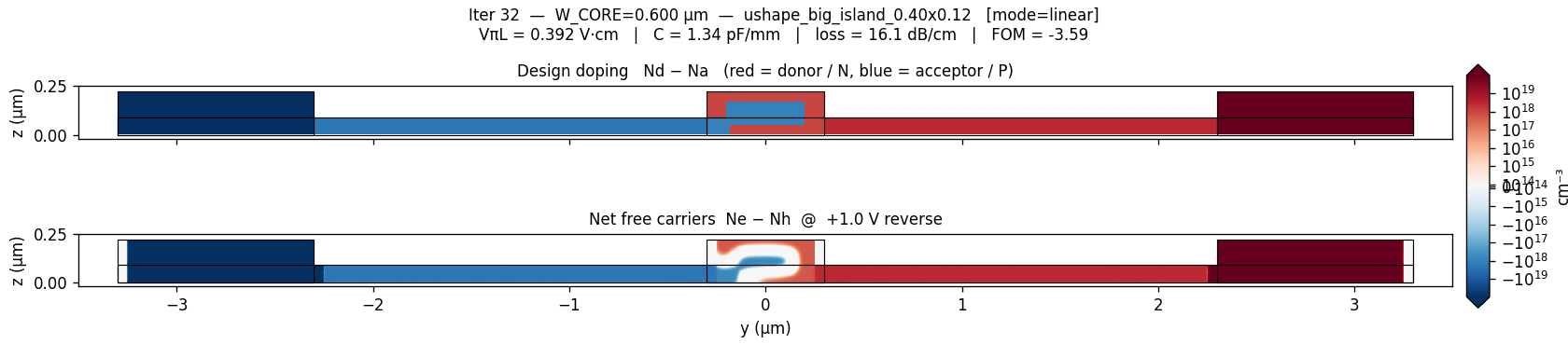}\\[2pt]\textbf{(c)}
\caption{\textbf{Active devices.} \textbf{(a)}~3-D view of the silicon
microring modulator with a lateral PN phase shifter. \textbf{(b)}~The
agent's 39 iterations in the $(\Cj, \VpiL)$ plane against the published
silicon-MRM compilation and its fit \citep{patel2026si}; the agent traces
the same envelope without being shown it, with iteration~28 (pink star) the
best $\VpiL\cdot\Cj$. \textbf{(c)}~Design doping and net free-carrier
distribution at \SI{1}{\volt} reverse bias for that design.}
\label{fig:active}
\end{figure}

Across 39 iterations the agent explored six junction-topology families and
traced a $\VpiL$-vs-$\Cj$ trade-off curve. The best design was a U-shape
junction with a \SI{300}{\nano\metre} $\times$ \SI{100}{\nano\metre} buried
P-island at $7 \times 10^{17}\,\mathrm{cm^{-3}}$ inside a symmetric
$7 \times 10^{17}\,\mathrm{cm^{-3}}$ N-outer (iterations 28/29;
Fig.~\ref{fig:active}), reaching
$\VpiL = \SI{0.53}{\volt\cdot\centi\metre}$ and
$\Cj = \SI{0.83}{\pico\farad\per\milli\metre}$ for a product
$\VpiL \cdot \Cj \approx \SI{4.4}{\volt\cdot\pico\farad}$. A generalized DRC
caught unfabricable doping stripes, electrically floating pockets, and
depletion regions that failed to overlap the optical mode. The agent reverted these designs
before any simulation ran so the design exploration stayed on physically
meaningful candidates (\S\ref{si:active}). The agent independently mapped
the same trade-off curve that a recent Nvidia OFC~2026 post-deadline
compilation assembled from a decade of published silicon modulators
\citep{patel2026si} (Fig.~\ref{fig:active}).

\subsection{RF electrode}
\label{sec:rf}

High-speed modulators encode information into light with a traveling microwave signal, but
their bandwidth is limited by the RF transmission-line electrode that carries it.
Therefore, accurate RF simulation and careful electrode design are central to modulator performance.
Designing transmission lines is typically a multi-objective problem with competing goals that are hard to satisfy at once.
Namely, it must achieve low microwave loss, velocity matching between the propagating
microwave and optical signals, and a characteristic impedance matched to the
source (typically \SI{50}{\ohm}) \citep{kharel2021breaking}.
Here, we encoded these competing objectives as one figure of merit and asked the agent to optimize a segmented slow-wave coplanar electrode (\S\ref{si:rf}).

\begin{figure}[H]
\centering
\begin{minipage}{0.46\textwidth}\centering
\includegraphics[width=\linewidth]{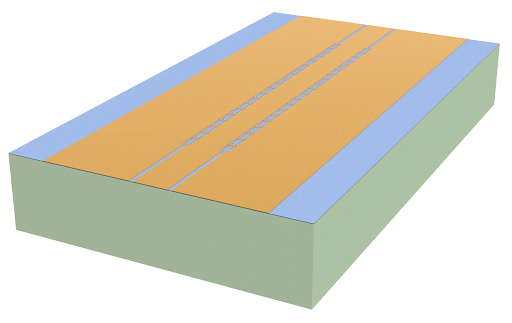}\\[2pt]\textbf{(a)}
\end{minipage}\hfill
\begin{minipage}{0.50\textwidth}\centering
\includegraphics[width=0.50\linewidth]{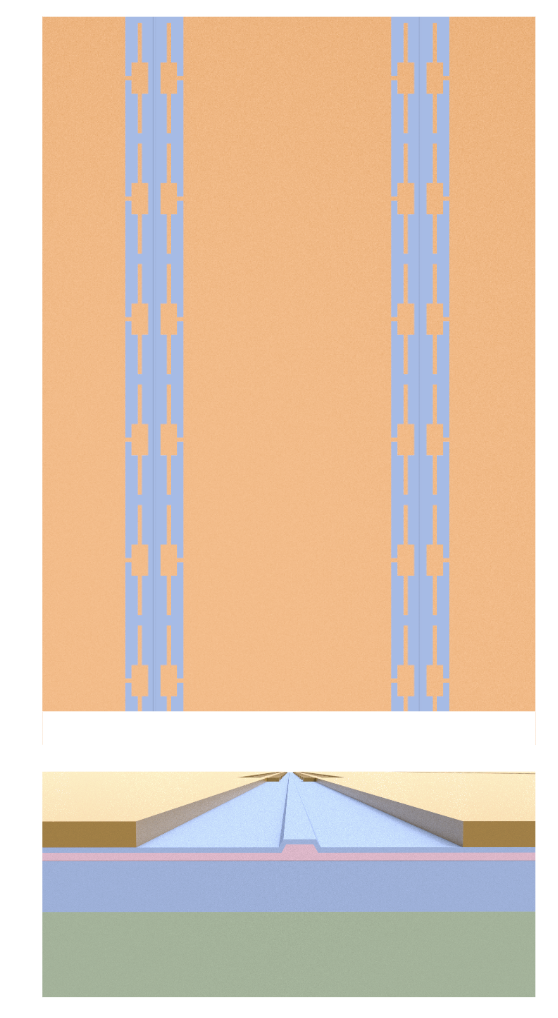}\\[2pt]\textbf{(b)}
\end{minipage}\\[0.6em]
\includegraphics[width=0.74\textwidth]{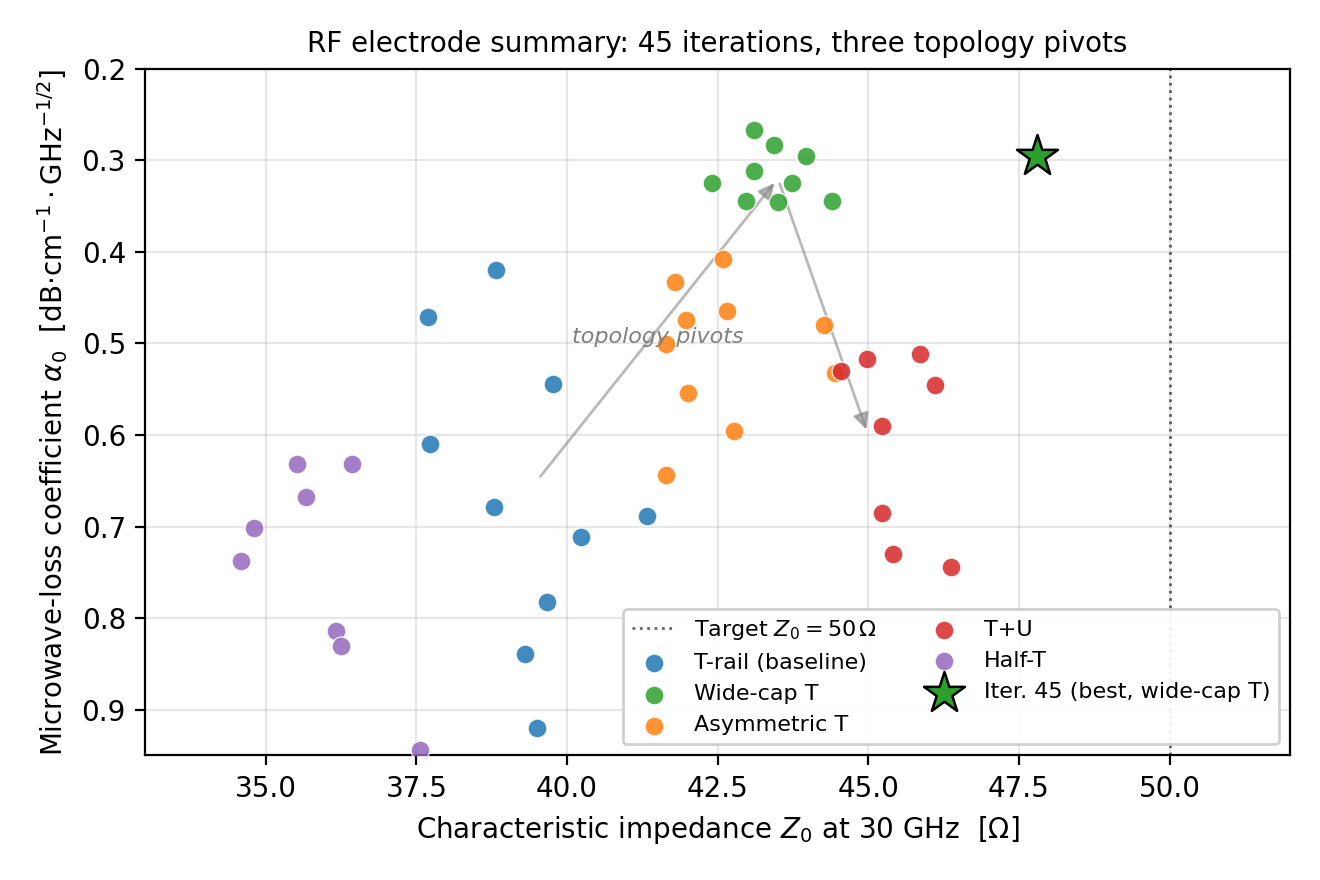}\\[2pt]\textbf{(c)}
\caption{\textbf{RF electrode.} \textbf{(a)}~3-D view of the segmented
coplanar electrode. \textbf{(b)}~Top-down and cross-section of one optimized
iteration. \textbf{(c)}~Run trajectory in the $(Z_0, \alpha_0)$ plane across
the five topology families; the wide-cap T family (best design marked)
clusters closest to the \SI{50}{\ohm}, low-loss, velocity-matched target.}
\label{fig:rf}
\end{figure}

Across 45 iterations the agent cycled between 5 electrode topologies, switching when its log showed the current one had plateaued in performance (Fig.~\ref{fig:rf}).
The agent logged a hypothesis before each change and corrected its assumptions against the simulation results as it went.
For example, the agent originally hypothesized that packing T-rails closer would raise capacitance.
Later, it was found that neighbouring rails shield one another, and the agent pivoted to lengthening individual rails instead.
The agent proposed a wide-cap T geometry that raised high-frequency capacitance while keeping series resistance roughly fixed, lowering microwave loss and pulling impedance toward the target.
The best design (iteration~45) reached $Z_0 \approx \SIrange{43}{44}{\ohm}$ and $\alpha_0 \approx 0.29$--$0.36\,\mathrm{dB\,cm^{-1}\,GHz^{-1/2}}$ over \SIrange{5}{45}{\giga\hertz}, versus $Z_0 \approx \SIrange{38}{41}{\ohm}$ and $\alpha_0 \approx 0.5$--$0.8$ on the plain T-rail baseline.

\subsection{Electrical routing}
\label{sec:routing}

External electrical signals need to be delivered to the various active
components placed within a chip through electrical routing. On a dense
photonic die this means routing dozens of nets at once without crossing
one another or touching sensitive layers like optical waveguides, while
obeying foundry spacing, width, and keepout rules. Manual routing is
tedious and slow, and even conventional autorouters handle the problem
only partially: the designer has a large set of degrees of freedom (the
choice of routing algorithm, the position of each net, the location of
each bondpad) that must be exercised while simultaneously satisfying the
geometric and fabrication constraints. We gave the agent a library of
twelve routing algorithms and the freedom to pick and switch among
them, and to shift bondpads within an allowed window.
The agent was instructed to write its own design-rule checks to flag traces crossing a heater layer or one another.
The layout was exposed to the agent through a low-order integer-grid representation, small enough that it could read the full state, see where violations occur, and reason about a fix before re-running the layout engine.

\begin{figure}[H]
\centering
\begin{tabular}{cc}
\includegraphics[width=0.46\textwidth]{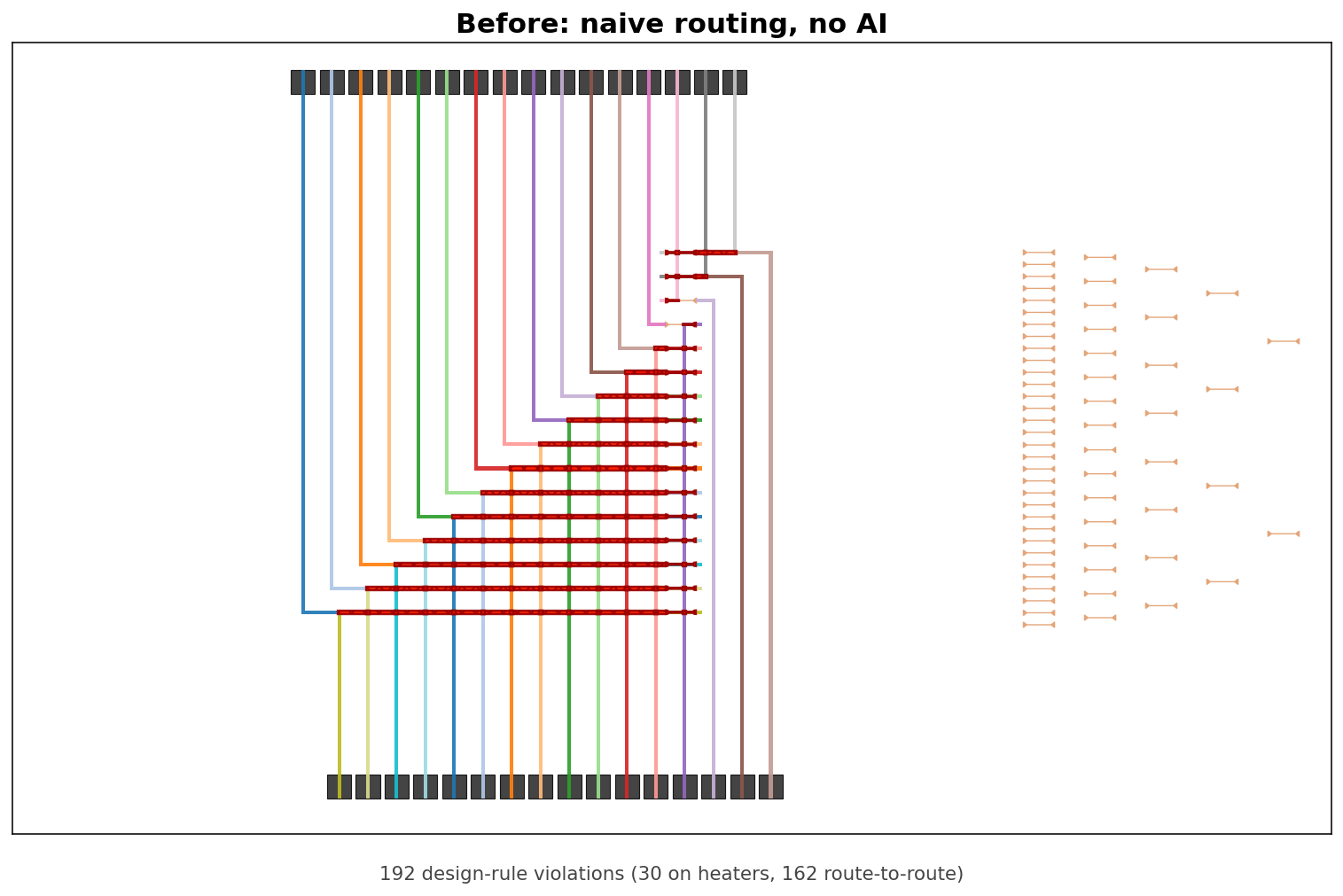} &
\includegraphics[width=0.46\textwidth]{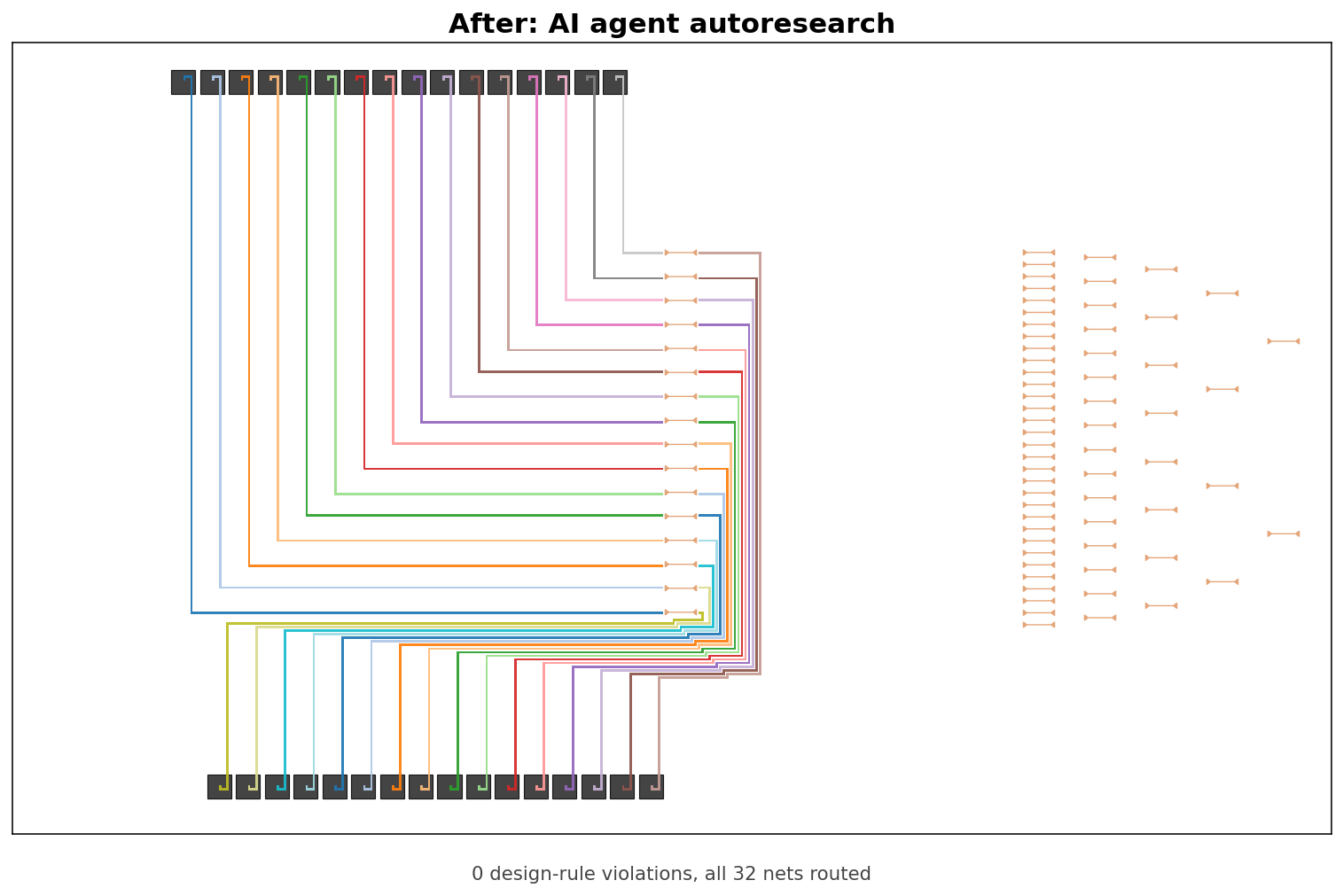} \\
\end{tabular}
\caption{\textbf{Electrical routing.} Starting layout (left): 192 DRC
violations from default per-pin autorouting. Final layout (right): zero
violations after 27 agentic iterations.}
\label{fig:routing_before_after}
\end{figure}

On a projector chip with 32 nets, we set up the initial routing problem
using a basic per-pin autorouter, which produced 192 DRC violations (30
heater cut-throughs and 162 route crossings). The agent then iterated
autonomously: it swapped the sequential per-pin router for a planner
that routes all 32 connections together on a shared grid, marked
heaters as inflated obstacles, reassigned pin-to-pad pairings to remove
unnecessary crossings, and widened the bondpad row to accommodate the
bundled traces. After 27 iterations it reached zero DRC violations
(Fig.~\ref{fig:routing_before_after}), in a total wall-clock time of
\SI{2}{minutes} \SI{25}{seconds}; a manual effort on the same chip
typically takes 2--3 hours (\S\ref{si:routing}). The same loop used to
design continuous device geometries was therefore also effective on a
discrete, combinatorial layout problem, with the agent guided by the
same kind of quantitative and verifiable performance signal as before,
here the number of DRC violations.

\subsection{End-to-end multiphysics modulator}
\label{sec:multiphysics}

In this section, we combine elements of all previous individual design
problems into a single demonstration of a photonic device that couples
all of these regimes. Here, we reproduce a published silicon photonic
modulator \citep{zhuang2024equivalent} end-to-end across layout, charge
transport, optical mode, and RF simulation.

In contrast to previous design problems, which were more
straightforward, setting up the initial simulation for this silicon
photonic modulator, which includes geometry, material stack, and doping
information, required some guidance from a human engineer.
\label{sec:multiphysics_reference}
While the agent set up the full multi-physics simulation autonomously
by referencing the paper, two setup issues required fixing before the
design loop could start: an initial mode mis-identification on the
loaded cross-section and a double-counted junction shunt. An engineer
corrected both mistakes in the setup stage.

\begin{figure}[H]
\centering
\begin{minipage}{0.49\textwidth}\centering
\includegraphics[width=\linewidth]{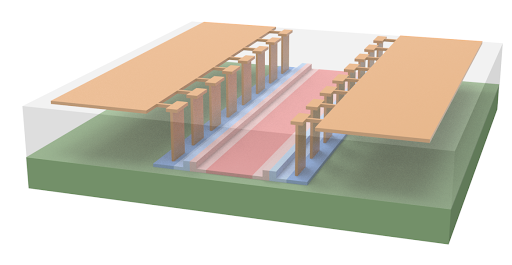}\\[2pt]\textbf{(a)}
\end{minipage}\hfill
\begin{minipage}{0.49\textwidth}\centering
\includegraphics[width=\linewidth]{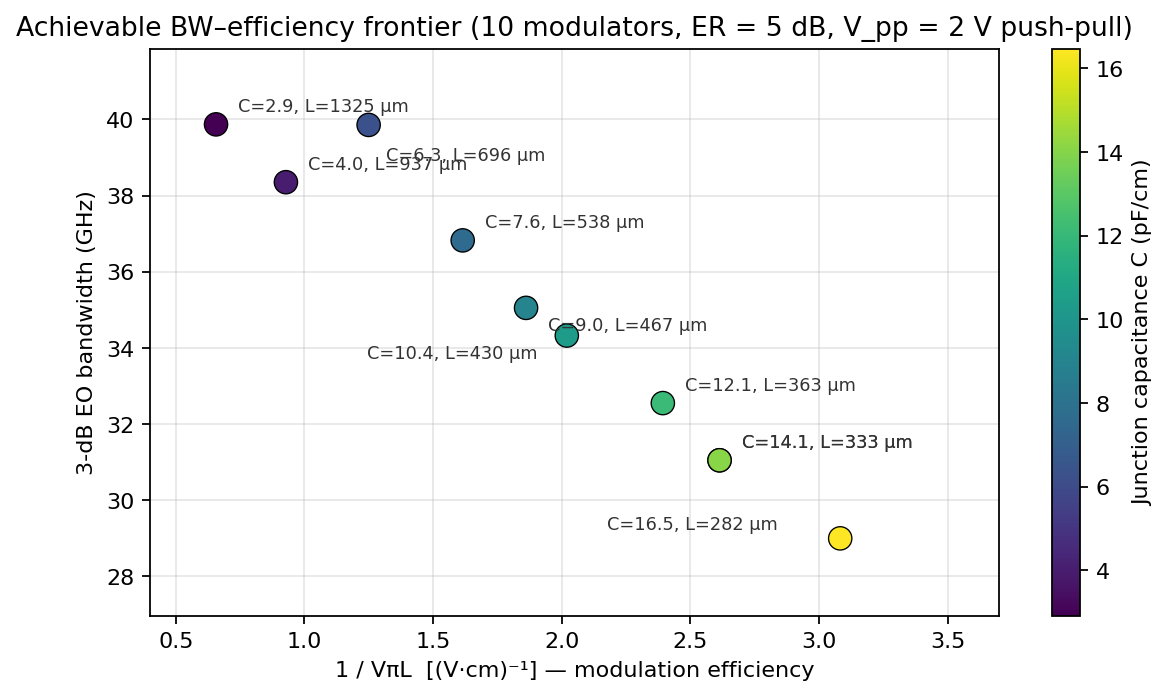}\\[2pt]\textbf{(b)}
\end{minipage}
\caption{\textbf{End-to-end multiphysics modulator.} \textbf{(a)}~3-D render
of the T-rail traveling-wave silicon Mach--Zehnder modulator reproduced in
this section. \textbf{(b)}~Bandwidth versus modulation efficiency for the
nine closed-loop designs; each is one fully optimized device, all holding
$Z_0$ within $\pm 10\%$ of \SI{50}{\ohm}.}
\label{fig:mzm}
\end{figure}

Starting from the corrected reference, the agent then ran the closed
design loop autonomously across the coupled charge, optical-mode, and
RF simulations. It edited geometry on each physics axis (junction
profile, electrode dimensions, segmentation period), kept or reverted
each candidate against the figure of merit, and journaled a hypothesis
before each move. Halfway through the run, the agent proposed and
adopted a modified objective that maximizes electro-optic bandwidth
directly rather than holding impedance and microwave index to fixed
targets. The loop produced nine distinct designs spanning the
bandwidth--efficiency trade-off curve (Fig.~\ref{fig:mzm}(b)), all
within $\pm 10\%$ of \SI{50}{\ohm} and spanning
\SIrange{29}{40}{\giga\hertz}.

\section{Discussion and conclusion}
\label{sec:discussion}
\label{sec:conclusion}

We have shown that a single agentic loop can coordinate components (passive,
active, RF), layout, and several physics domains (charge, optical, and RF)
to autonomously design photonic devices. The success of this demonstration relied on two core capabilities:
a) programmable simulation and evaluation tools (Python-scriptable
solvers spanning the relevant physics, with design-rule checks the agent can
call directly) and b) frontier coding agents capable enough at code and
reasoning to drive the loop end-to-end.

Performance was further enhanced by a model context protocol (MCP) interface
for tighter agent integration with the simulation tools (\Tidy,
\PhotonForge), an in-house GPU cluster that lets the agent run hundreds of
simulations per design problem in reasonable time, and a workflow that guides the agent through literature review, hypothesis generation, and review of previous iterations.

This approach generalizes previous methods for automated photonic design,
which were commonly driven by gradient-based optimization (``inverse
design'' in the photonics community \citep{molesky2018inverse, hughes2018adjoint, shi2026adjoint}). Agent-driven design is
complementary to these approaches: the agent can invoke gradient-based
inverse-design optimizers as one tool inside the loop, and did so in several
of the passive-component runs. It is also free to propose new candidate
parameterizations and handle non-differentiable choices such as routing,
material selection, and budget-aware stopping rules.

While the design loop itself is autonomous, an engineer is still required to supply
the acceptance and evaluation criteria and tune the initial setup. A natural
next step is to migrate more of the problem initialization to the agent,
including benchmarking, simulation setup, and the definition of performance
criteria. In addition, incorporating corrections to designs based on
previous fabrication and measurement results into this autonomous loop
would help close the loop from design to measurement.
In that framework, the engineer's role shifts from supervising
execution to supervising intent. For many simple problems, such as passive
component design, this is already within reach. For complex, highly coupled
multiphysics problems such as the MZM demonstrated here, an expert engineer
is still required to craft the problem definition before execution can be
fully automated.

\section{Code and reproducibility}
\label{sec:code}

All code for the demonstrations in this paper is available on GitHub. The
individual-component workflows live in
\href{https://github.com/flexcompute/autophotonicdesign}{\texttt{flexcompute/autophotonicdesign}};
the multiphysics MZM demonstration lives in
\href{https://github.com/aminkhavasi/modulator-autodesign}{\texttt{aminkhavasi/modulator-autodesign}}.

All demonstrations were run with Anthropic's Claude Opus 4.7 in Max Effort
mode unless otherwise noted.

\section*{Author contributions}

P.K.\ developed the demonstrations and led the manuscript revisions; primary
responsibility for \S\ref{sec:active}--\ref{sec:routing} (active modulator,
RF electrode, and electrical routing).
A.K.\ developed the end-to-end multiphysics modulator demonstration
(\S\ref{sec:multiphysics}).
X.C.\ contributed to the passive-component demonstrations
(\S\ref{sec:passive}) and prepared figures.
T.W.H.\ conceived the project, coordinated the work, and wrote and edited
the manuscript.

\clearpage

\clearpage

\appendix
\renewcommand{\thesection}{S\arabic{section}}
\renewcommand{\thefigure}{S\arabic{figure}}
\renewcommand{\thetable}{S\arabic{table}}
\setcounter{figure}{0}
\setcounter{table}{0}

\begin{center}
\vspace*{1em}
{\LARGE\bfseries Supplementary Information}\\[0.5em]
{\large Autonomous agentic design for photonics}
\end{center}
\vspace{1.5em}

\section{Positioning against prior and adjacent work}
\label{si:positioning}

\begin{table}[!htbp]
\centering
\small
\setlength{\tabcolsep}{6pt}
\caption{Positioning of this work against representative prior and adjacent
work in agentic and classical PIC design automation, on five axes.
\checkmark{} = clearly demonstrated; \emph{partial} = partially demonstrated;
-- = not demonstrated. Footnotes clarify selected entries.}
\label{tab:positioning}
\begin{tabular}{lccccc}
\toprule
 & Outer & Multi- & Multiple & Chip & Reproduces \\
Work & EM & physics & device & layout & published \\
 & loop & & classes & / routing & device \\
\midrule
Agentic metasurface design \citep{lupoiu2025metachat, huang2026selfevolving, wu2026metadesigner, lu2025agentic, kim2025nanophotonic} & \checkmark$^{a}$ & -- & -- & -- & -- \\
\citet{sharma2025aiagents} (PhIDO) & --$^{b}$ & -- & partial & partial & -- \\
\citet{meng2026fermilink} (generalist) & partial & partial & \checkmark & -- & partial$^{c}$ \\
\citet{zhou2025polaris} (no LLM) & -- & -- & partial & \checkmark & -- \\
\textbf{This work} & \checkmark & \checkmark & \checkmark & \checkmark & \checkmark$^{d}$ \\
\bottomrule
\end{tabular}\par
\smallskip
\footnotesize\raggedright
$^{a}$Single inverse-design family (metasurfaces/metamaterials); the outer
loop is closed on full-wave or differentiable EM feedback.
$^{b}$Verification via S-parameter netlists; an FDTD loop is closed on one
device in a public branch.
$^{c}$Reproduces published figures across $\sim$50 packages, not a coupled
device workflow.
$^{d}$Full end-to-end reproduction of \citet{zhuang2024equivalent}; the
engineer iterated on the initial problem setup.
\end{table}

\section{Passive generalization across device classes (\S\ref{sec:passive})}
\label{si:passive}

\begin{figure}[!htbp]
\centering
\includegraphics[width=\textwidth]{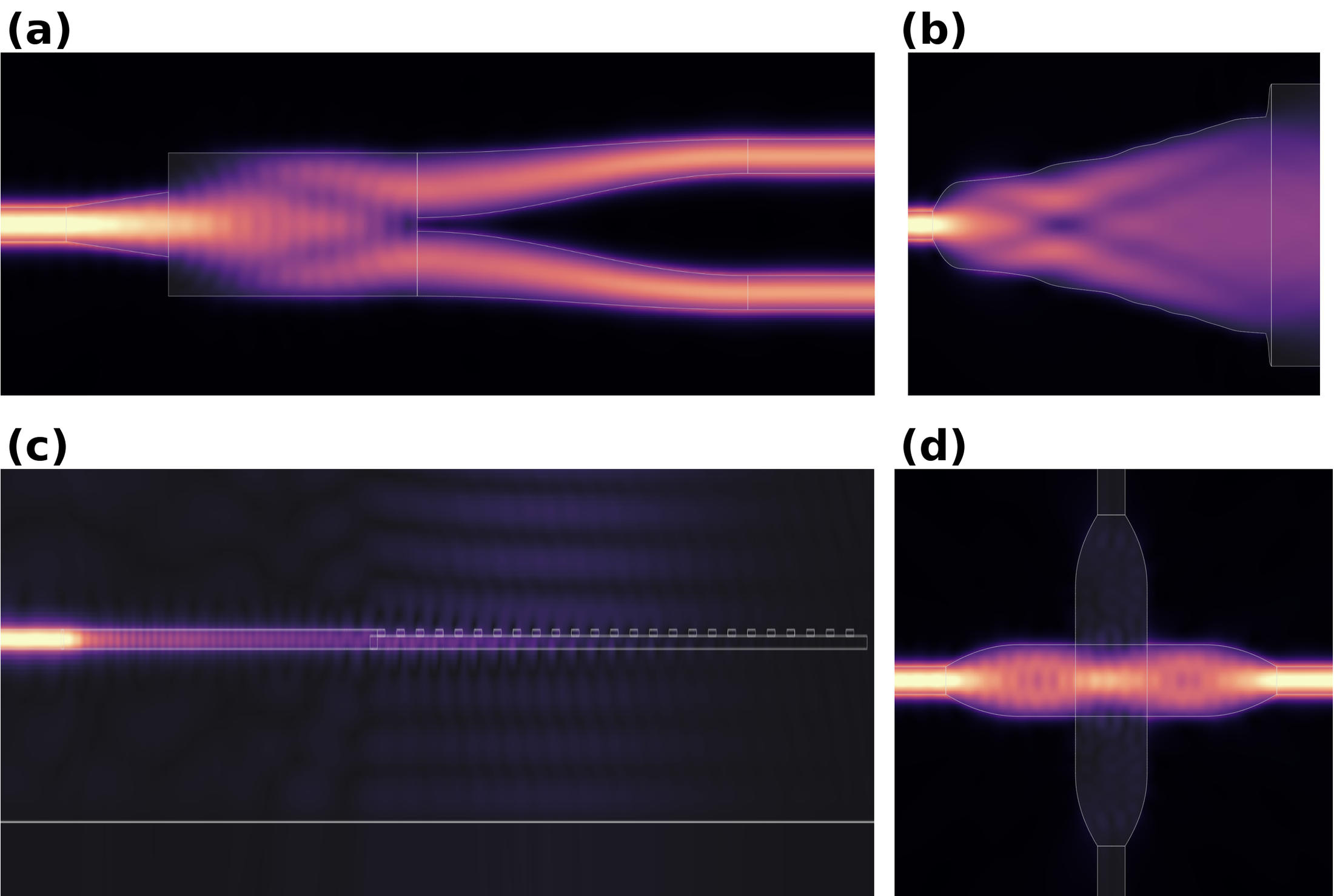}
\caption{Final steady-state fields from the four
\SI{220}{\nano\metre} SOI passive runs at \SI{1550}{\nano\metre}:
\textbf{(a)} 1$\times$2 MMI splitter (\SI{0.015}{\decibel});
\textbf{(b)} 0.5--5\,$\mu$m taper over \SI{6}{\micro\metre}
(\SI{0.114}{\decibel}); \textbf{(c)} partially etched focused grating
coupler (\SI{2.89}{\decibel} peak coupling loss); and \textbf{(d)} TE crossing
in a \SI{6}{\micro\metre}$\times$\SI{6}{\micro\metre} design region
(\SI{0.11}{\decibel}).}
\label{fig:passive_others}
\end{figure}

We reused the waveguide bend workflow for four additional passive devices
in a \SI{220}{\nano\metre} silicon-on-insulator (SOI) platform at C-band: a
compact 1$\times$2 splitter, a short waveguide taper, a partially etched
focused grating coupler, and a waveguide crossing. The agent reached
\SI{0.015}{\decibel} fundamental-mode insertion loss for the splitter,
\SI{0.114}{\decibel} for a \SI{6}{\micro\metre} 0.5--5\,$\mu$m taper,
\SI{2.89}{\decibel} peak coupling loss for the grating coupler, and
\SI{0.11}{\decibel} insertion loss for a
\SI{6}{\micro\metre}$\times$\SI{6}{\micro\metre} crossing
(Fig.~\ref{fig:passive_others}).

\paragraph{Platform and figure of merit.}
We instantiate the workflow of \S\ref{sec:workflow} with a
\SI{400}{\nano\metre} SiN core on SiO$_2$ cladding, a
\SI{1.2}{\micro\metre} reference waveguide width, and a
\SI{150}{\nano\metre} minimum-feature-size fabrication constraint.
The figure of merit was the fundamental-mode transmission at \SI{1550}{\nano\metre}.

\subsection{Results}
\label{sec:passive_results}

The agent was given a budget of 50 simulations per design and a circular bend baseline.
The design optimization trajectory is shown in
Figure~\ref{fig:bend}(a).

The final design was a single parametric bend combining three geometric
modifications. First, the circular baseline was replaced by a
Euler-circular-Euler bend with Euler fraction $p_E = 0.45$. The two Euler
sections occupied \SI{40.5}{\degree} of the \SI{90}{\degree} sweep, and the
remaining \SI{49.5}{\degree} was circular. The curvature ramped
linearly through each Euler section, so $\kappa$ was continuous at the joins with
both the straight I/O guides and the central arc, removing the junction
curvature step that drives mode-mismatch loss. Second, the local waveguide
width was modulated as
\begin{equation*}
  w(u) = w_0\left[1 + (w_r - 1)\sin^2(\pi u)\right],
\end{equation*}
with $w_0 = \SI{1.2}{\micro\metre}$ and $w_r = 2.0$, so the guide widened
smoothly to \SI{2.4}{\micro\metre} where curvature was largest and returned to
$w_0$ at each endpoint. This improved bent-mode confinement without creating
an I/O width mismatch. Third, the centerline was displaced perpendicular to
its local tangent by
\begin{equation*}
  \Delta r(u) = r_0\sin^2(\pi u),
\end{equation*}
with $r_0 = \SI{-1.1}{\micro\metre}$, an inward bow toward the bend center
that peaked at the midpoint. The bow relaxed the average effective curvature
while concentrating the tightest bend near the midpoint, where the mode had
already adapted. The shared $\sin^2(\pi u)$ envelope matched the boundary on both ends:
its value and derivative vanished at $u = 0$ and $u = 1$, so the width and
centerline met the straight I/O waveguides without a kink or width step.

The agent first replaced the circular arc with
an Euler arc, lifting transmission from \SI{89.96}{\percent} to
\SI{92.73}{\percent}. After an outward centerline shift failed in
simulation, it inverted the shift to an inward bow and reached
\SI{96.02}{\percent}. Tuning the width taper and the offset together
then produced the best design at experiment~38.

As shown in Figure~\ref{fig:bend}(b), the final design achieved
\SI{97.51}{\percent} fundamental mode transmission
(\SI{0.109}{\decibel} bend loss) at \SI{1550}{\nano\metre}. Compared with
\SI{89.96}{\percent} (\SI{0.460}{\decibel}) for the circular baseline, it was a
roughly fourfold loss reduction at fixed radius. The inset field profile in
Figure~\ref{fig:bend}(b) shows the mode tracking the outer wall of the
widened section with negligible inner-sidewall radiation.

For a \SI{1.2}{\micro\metre} by \SI{0.4}{\micro\metre} SiN waveguide at a
\SI{12}{\micro\metre} bend radius, \SI{0.11}{\decibel} falls in the same
low-loss regime as published bend
studies \citep{cherchi2013dramatic, vogelbacher2019analysis,
bahadori2019universal}. We do not claim a one-to-one benchmark ranking,
because stack details, mode definitions, wavelengths, and fabrication
constraints differ across the literature. The agent reached this regime
from a circular-bend baseline without per-iteration human intervention.

\section{Active device: layout and per-bias cross-section (\S\ref{sec:active})}
\label{si:active}

\begin{figure}[!htbp]
\centering
\includegraphics[width=0.98\textwidth]{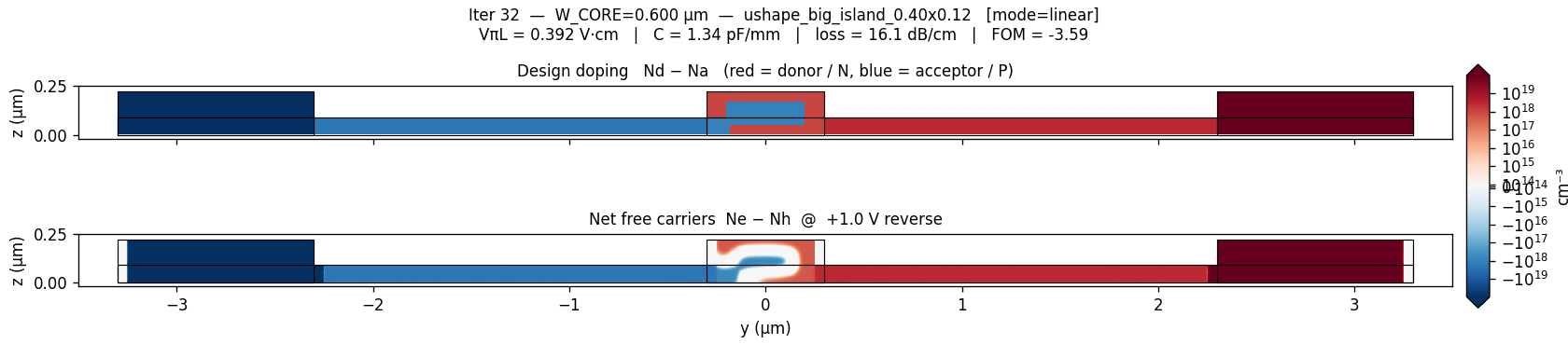}
\caption{One cross-section from the run (iteration 32, U-shape with a
buried P-island): design doping $N_d - N_a$ (upper) and net
free-carrier distribution $N_e - N_h$ at \SI{1}{\volt} reverse bias
(lower, white = depletion). The run sweeps 39 such profiles across six
junction-topology families.}
\label{fig:mrm_overview}
\end{figure}


\subsection{Loop adaptations}

We reused the same workflow from \S\ref{sec:workflow}. Three additions were included in this design problem.

\paragraph{Editable design space.}
The agent's parameters were a list of doping regions (each with
concentration, geometric extent, and topology family) and a rib-width
parameter. The agent could add, remove, or reshape doping regions and
switch between topology families freely.

\paragraph{Generalized DRC.}
Layout DRC normally checks geometry, such as minimum widths, spacings, and acute
angles. Here, we allowed the agent to check additional criteria. Doping stripes below \SI{100}{\nano\metre} were
flagged as unfabricable; doping pockets isolated from any contact rail
were flagged as electrically floating; and depletion regions that failed
to overlap the optical mode at the operating bias were flagged as
electro-optically inactive. A failed check stopped the iteration before
any simulation was run to keep the simulation budget reserved for geometries
that were physically meaningful.
\paragraph{Simulation at multiple reverse bias points.}
Each evaluation ran \Tidy's steady-state drift-diffusion charge solver at five reverse-bias points, then solved the optical mode at each bias with a carrier-perturbation medium \citep{soref1987electrooptical}, yielding $\VpiL(V)$ and $\Cj(V)$.
The agent sought to minimize the product $\VpiL \cdot \Cj$ at the operating bias, which is the figure of merit for modulator energy-per-bit.
Propagation loss and DRC violations entered as penalties so that fabrication and insertion-loss constraints stayed in scope.

\subsection{Results}
Across 39 iterations the agent explored six junction-topologies and produced a set of designs that traced out a $\VpiL$-vs-$\Cj$ trade-off curve.
The best design was a U-shape junction with a $300 \times \SI{100}{\nano\metre}$ P-island buried \SI{50}{\nano\metre} above the slab inside a $7 \times 10^{17}\,\mathrm{cm^{-3}}$ N-outer.
At \SI{1}{\volt} reverse bias the depletion region wrapped around the buried P-island and overlapped the peak of the optical mode (device layout and biased cross-section in \S\ref{si:active}).
$\Cj$ dropped monotonically as the depletion widened, and $\VpiL$ rose as the widening pushed free carriers out of the optical mode.
Figure~\ref{fig:active} plots the run in the $(\Cj, \VpiL)$ plane.
The same trade-off curve was independently mapped by Nvidia in a recent OFC 2026 post-deadline compilation that aggregated a decade of published silicon MRMs across six junction-topology families \citep{patel2026si}.
The agent's clusters landed along that curve without having been shown the envelope, the fit, or any of the published designs.
\subsection{Discussion}
This exploration successfully identified where in the topology space to invest deeper modelling and process work.
A natural next step would be to couple the agent to the process tools that decide whether a doping profile is actually manufacturable.
The carrier distributions used here were idealized boxes at chosen doping densities and depths; a real implant has straggle, lateral diffusion, and anneal-driven redistribution that the device-level solver does not capture.
Letting the agent also drive a TCAD process simulator, with the implant recipe in its editable design space, would let the loop optimize for doping topologies that minimize $\VpiL \cdot \Cj$ and are reachable on the foundry process.


\section{RF electrode: 3-D geometry and topology families (\S\ref{sec:rf})}
\label{si:rf}


\subsection{Loop adaptations}

In this design study, we re-used the workflow from \S\ref{sec:workflow}.
The new pieces were the simulator (a 3-D full-wave RF solver), the figure
of merit, and the parameterization.

\paragraph{Editable design space.}
We allowed the agent to change the T-rail geometry
(along-propagation period, transverse extent, neck length and width,
cell-to-cell spacing), the host coplanar waveguide (CPW) (signal trace
width, ground rail width, residual gap above the rib waveguide), and the
metal stack (Au thickness, cladding gap above the slab). We further
grouped the geometries into topology classes that the agent could swap
between freely. Thin-film thicknesses, sidewall angle, frequency band,
and substrate permittivities were fixed by the platform.

\paragraph{Topology families.}
The agent had the freedom to choose between different transmission-line
topologies: symmetric T-rail (the published baseline), asymmetric T
(independent signal- and ground-side neck widths), wide-cap T (a broad
capacitive hat on each T-top), T+U (T-rails interleaved with U-shaped
signal-bridges every $N$th cell), and half-T (signal-anchored only).

\paragraph{Generalized DRC.}
As in \S\ref{sec:active}, DRC included both fabrication rules
(\SI{100}{\nano\metre} minimum metal feature, spacing, and thickness;
residual ground-rail width $\geq \SI{100}{\nano\metre}$ after T-rail
loading) and process constraints (thin-film thicknesses fixed; cladding
gap monotonically non-decreasing).
\paragraph{Figure of merit.}
During each evaluation, the agent ran a 3-D full-wave RF simulation
using \Tidy{} over \SIrange{5}{45}{\giga\hertz}, fit
$\alpha(f) \approx \alpha_0 \sqrt{f}$, and extracted $Z_0$ and $\neff$
from the dominant mode. The figure of merit combined the three
properties of the electrode: velocity matching ($\neff$ at the optical
group index $2.20$ at \SI{1310}{\nano\metre}), impedance matching ($Z_0$
at \SI{50}{\ohm}), and low microwave loss ($\alpha_0$ as small as
possible). The scalar objective was

\begin{equation}
\mathrm{FOM} = -\!\left[\,\alpha_0
+ \lambda_Z\!\left(\frac{Z_0 - \SI{50}{\ohm}}{\SI{50}{\ohm}}\right)^{\!2}
+ \lambda_n\!\left(\frac{\neff - 2.20}{2.20}\right)^{\!2}\right],
\label{eq:fom_rf}
\end{equation}

with $\lambda_Z = 5$ and $\lambda_n = 50$.
The two quadratic terms penalize deviations from the $Z_0$ and $\neff$ targets and turn off when those targets are hit.
The $\alpha_0$ term was uncapped so the agent always sees a benefit from lower microwave loss.

\subsection{Results}
Across 45 iterations the agent worked through five topology families,
recording each pivot as a hypothesis in its own journal. The narrative
below summarizes what the agent journaled during the autonomous run. T+U
and half-T variants were explored but neither outperformed the best
wide-cap T design.

In the first phase, the agent swept the symmetric T-rail baseline. It
discovered that increasing the transverse length of the T-rail (relative
to the propagation axis) raised $\neff$ super-linearly, because the
parallel-plate area between the T-top and the high-permittivity substrate
dominates per-cell capacitance. Once it identified this transverse length
as the primary knob for the microwave index, the agent started testing
several hypotheses around it. It found that shrinking the inter-T spacing
along the propagation direction \emph{reduced} per-cell capacitance,
because neighbouring T-rails coupled their fields and partially shielded
one another; the agent therefore kept lengthening each T-rail
transversely rather than packing them closer along propagation. It also
found that widening the host CPW gap over the rib waveguide
\emph{lowered} microwave loss, because the inductive path through the
host CPW dominated and a wider gap dropped host resistance per unit
length.

The agent then pivoted to asymmetric T-rails and found that the
capacitive load was gated by the shorter of the two T-rails, so for a
fixed total metal area the asymmetric variant carried lower per-cell $C$.
This led the agent to reframe asymmetric loading as a
$Z_0$/$\alpha_0$ tuning knob rather than the $\neff$ knob it had first
hypothesized.

For its third pivot, the agent moved to wide-cap T (a broad capacitive
hat on each T-top), which produced the best FOM in the run. The hat saw
the field but did not carry the bulk signal current, so it raised
high-frequency capacitance without proportionally raising $R$. On the
plain T-rail baseline the agent had been stuck at
$\alpha_0 \approx 0.5$--$0.8\,\mathrm{dB \cdot cm^{-1} \cdot
GHz^{-1/2}}$ with $Z_0 \approx \SI{38}{\ohm}$--\SI{41}{\ohm}; the
wide-cap T variant brought $\alpha_0$ down to $\approx 0.29$--$0.36$ and
pulled $Z_0$ up to $\approx \SI{43}{\ohm}$--\SI{44}{\ohm}, while the
asymmetric handles supplied the residual $\neff$ tune. The agent's best
design (iteration 45) was a wide-cap T with a \SI{53}{\micro\metre}
transverse T-rail, a \SI{15}{\micro\metre}$\times$\SI{8}{\micro\metre}
cap on top, and a \SI{12}{\micro\metre} host CPW gap, with all other
parameters at the platform default.

\subsection{Discussion}

This run showed two patterns that did not appear in the previous two
sections.

\paragraph{Topology pivots from journal evidence.}
The 45-iteration trajectory was not a sweep over a fixed parameterization.
It was an interleaved exploration of five geometry families. The agent
stayed inside a family while the FOM kept improving and pivoted when
its own journal entries showed the family had plateaued. The run
recorded three such pivots, and each one carried a journal entry naming
the metric that had stalled and the mechanism the new family was
expected to address (for example, \textit{wide-cap T to add $C$ without
proportionally raising $R$}). None of these pivots were scripted; the
agent made them by reading its own log.

\paragraph{Agent-initiated infrastructure fixes.}
During the run, the agent identified and corrected three issues in the
simulation infrastructure that we had not anticipated. The wave-port
domain became too small for short-period geometries, and the agent
patched it by adjusting the pad-length rule. A builder bug caused
adjacent T-tops to overlap at certain parameter combinations, and the
agent fixed it by recomputing the period. A band-edge artefact
contaminated the conductor-loss extraction, and the agent fixed it by
tightening the fit window. The agent diagnosed each of these by reading
its own simulation logs, noticing a discrepancy between expected and
observed physics, and editing the infrastructure itself.

\paragraph{Platform portability.}
The same scaffold transfers to other modulator platforms by changing
the material stack, the FOM targets, and the topology builders. The
simulation infrastructure, DRC, and journal carry across unchanged.
Production-grade modulator electrodes typically involve multiple metal
layers, dedicated VIA structures, ground straps, and substrate undercut
or defected ground structures, with 30--50 free parameters per geometry.
Manual exploration converges slowly at that dimensionality; the agentic
loop tolerates it because the cost per iteration is the cloud simulation,
not the design decision.


\section{Routing: DRC trajectory (\S\ref{sec:routing})}
\label{si:routing}

\begin{figure}[!htbp]
\centering
\includegraphics[width=0.86\textwidth]{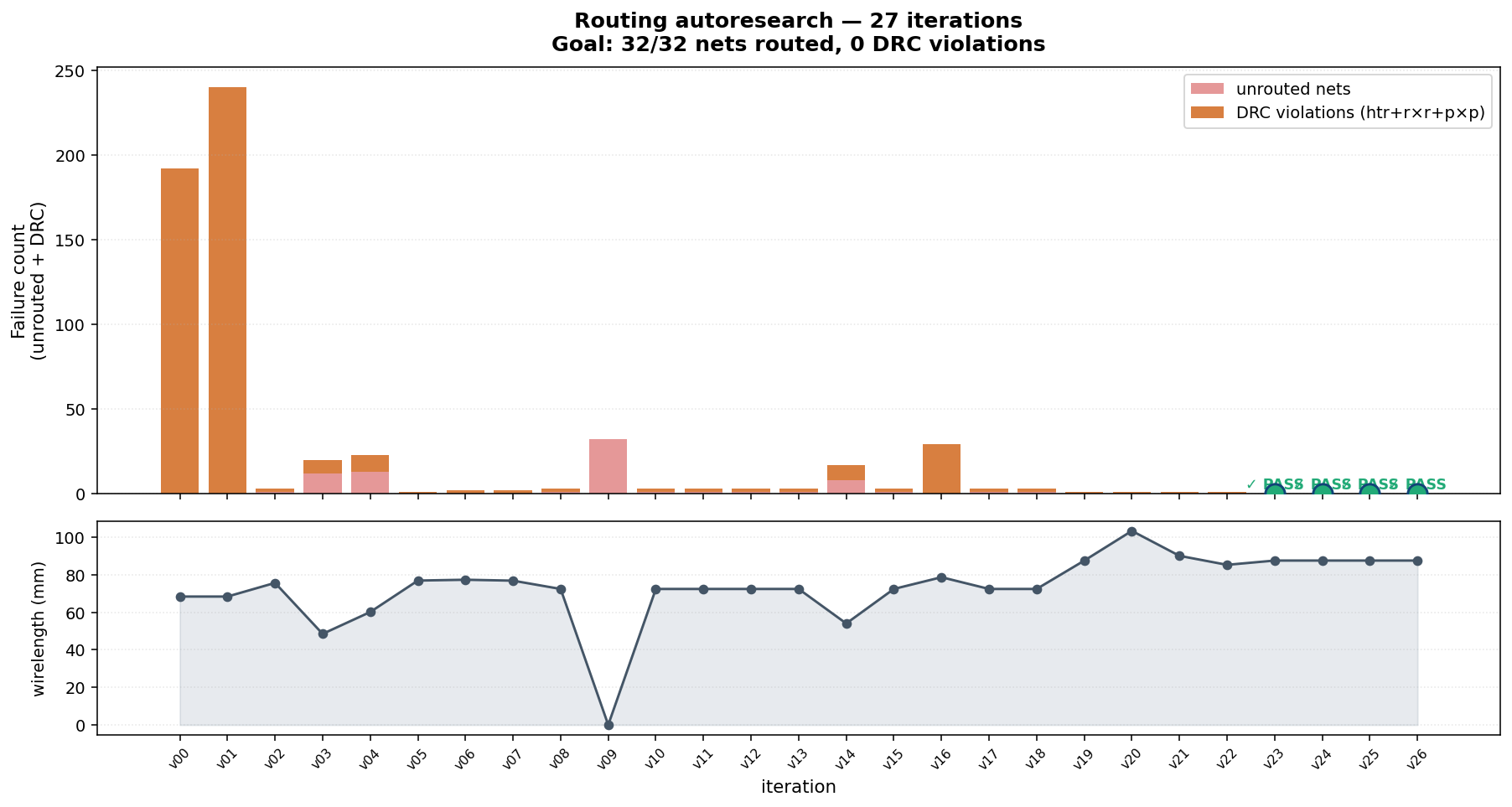}
\caption{DRC violation count vs.\ iteration. 192 violations at
iteration~0 drop to zero at iteration~27. Route-to-route crossings and
heater cut-throughs are tracked separately so the agent's journal can
target whichever class is binding.}
\label{fig:routing_loss}
\end{figure}


\subsection{Loop adaptations}
In this design study, we re-used the workflow from \S\ref{sec:workflow},
swapping the device simulator for \PhotonForge\ as the layout backend
(\PhotonForge\ does many other things besides layout; here we used only
its layout API). We built a verification stack that discretized the chip
onto a routing grid, detected route-to-route collisions, flagged heater
crossings and pad overlaps, and returned a single DRC violation count
along with per-class counts so the agent's journal could track which
constraint type drove each iteration.

We gave the agent the same freedom a human router would have: it could
edit a routing script, pick from a library of routing algorithms
(Manhattan, grid BFS, A*, bundle routing, rip-up-and-reroute, capacity
mesh, negotiated rip-up), shift bondpads within an allowed window,
reassign pin-to-pad pairings, and re-run DRC after every change.

A practical lesson from this study was that the agent worked much faster
when we exposed the layout as an integer grid, with each cell tagged as
wire, obstacle, or free. The grid was small enough that the agent could
read the full state, point to the cells where violations occurred, and
reason about a move before re-running \PhotonForge. Iterating on raw GDS
polygons would not have given the same feedback loop.

\subsection{Results}
Our test case was a projector chip with 32 metal contact points that had
to reach 32 bondpads at the die edge. The starting script was the
default: \PhotonForge's built-in point-to-point router called once per
pin in isolation, with no shared routing fabric. This produced
\textbf{192 DRC violations}: 30 routes cut through a heater and 162
route-pairs crossed one another (Fig.~\ref{fig:routing_before_after},
left).

Across \textbf{27 iterations the agent reduced the violation count to
zero} (Fig.~\ref{fig:routing_loss}). The sequence of moves the agent
made was: switching from independent point-to-point routing to a
grid-based planner so that nets shared the fabric; marking heaters as
inflated obstacles to force routes around them; revising the pin-to-pad
assignment to remove unnecessary crossings; widening the bondpad row to
accommodate the bundled traces; and adding a local constraint for a
single contact whose proximity to a heater the bulk planner could not
resolve. The agent logged each move with the violation class it targeted
and the resulting count.

The total wall-clock time for the 27-iteration run, including every
failed attempt, was \textbf{2 minutes 25 seconds}. A human engineer on
our team typically spends 2--3 hours to reach a DRC-clean result on this
chip; we use that as an internal baseline rather than a controlled
benchmark.

\subsection{Discussion}
The same agentic loop we used in
\S\S\ref{sec:passive}--\ref{sec:rf} transferred to routing without
modification. Routing is a verifiable domain: the DRC stack produces a
deterministic scalar, the layout is a deterministic function of the
script, and the agent can keep or revert any move based on the
violation count. The agent chose which routing algorithm to use, when
to switch, how to assign pins to pads, and where to place obstacles.
The library exposed the primitives; the agent assembled them into a
strategy by reading the violation breakdown after each iteration. None
of the moves in the final layout (grid planning, obstacle inflation,
pin-to-pad reassignment, bundle widening, the single-contact exception)
were prescribed in the loop instructions.


\section{Multiphysics MZM: stage-by-stage reproduction (\S\ref{sec:multiphysics})}
\label{si:multiphysics}

This section walks through the multiphysics MZM reproduction in three parts: the first attempt where the agent set up the full pipeline but failed to match the RF metrics; the human-expert reference that fixed the RF stage; and the closed-loop optimization run that followed. Full run journals are in the released repositories (\S\ref{sec:code}).

\begin{figure}[!htbp]
\centering
\includegraphics[width=0.85\textwidth]{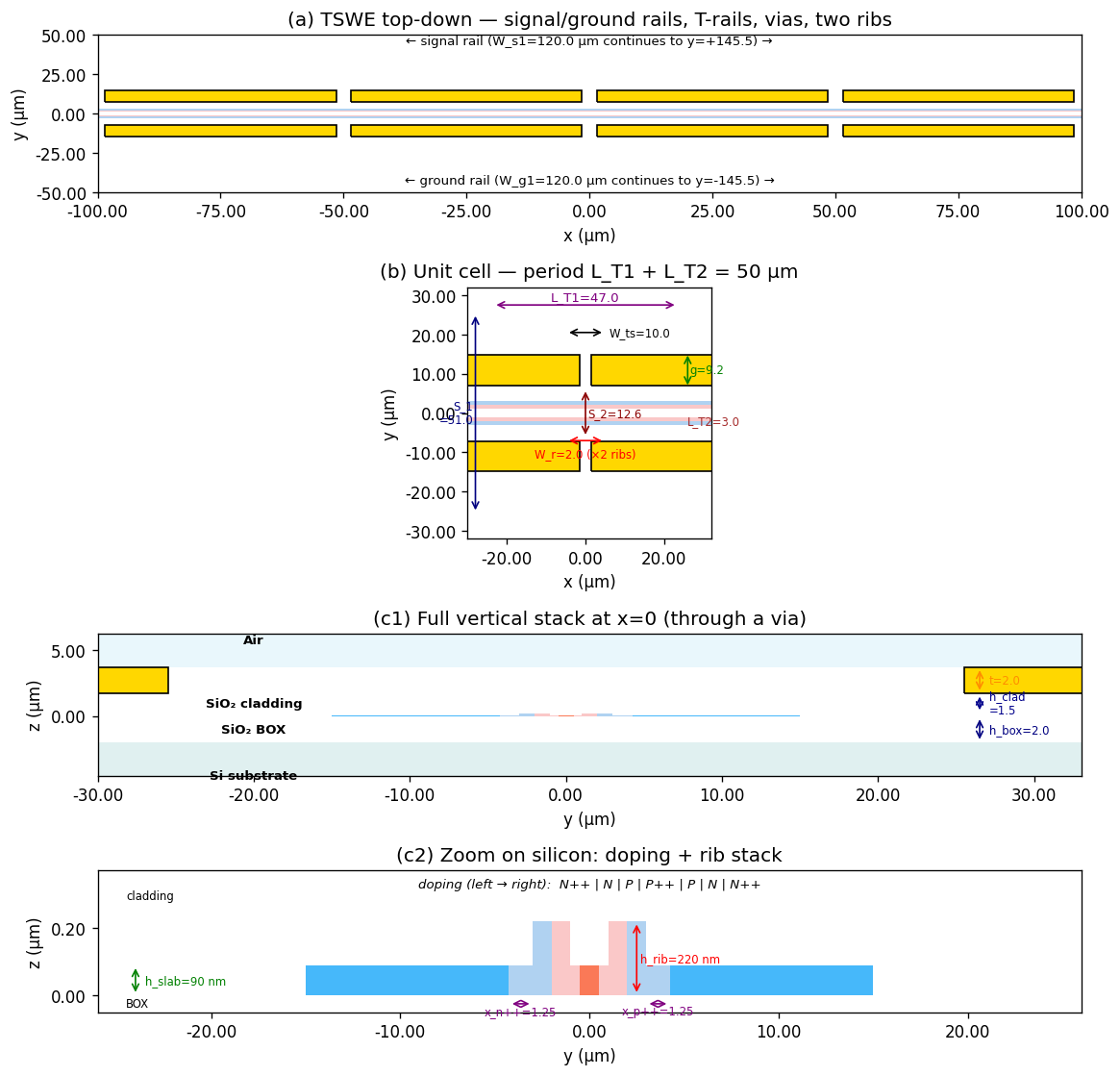}
\caption{Agent-built parametric layout of the Zhuang T-rail MZM.
\textbf{(a)}~Top view of the T-loaded CPS with two push--pull ribs.
\textbf{(b)}~Annotated 50\,$\mu$m unit cell. \textbf{(c)}~Vertical
stack and rib cross-section with the geometric and doping parameters
matched to the reference.}
\label{fig:mzm_layout}
\end{figure}

\begin{figure}[!htbp]
\centering
\includegraphics[width=\textwidth]{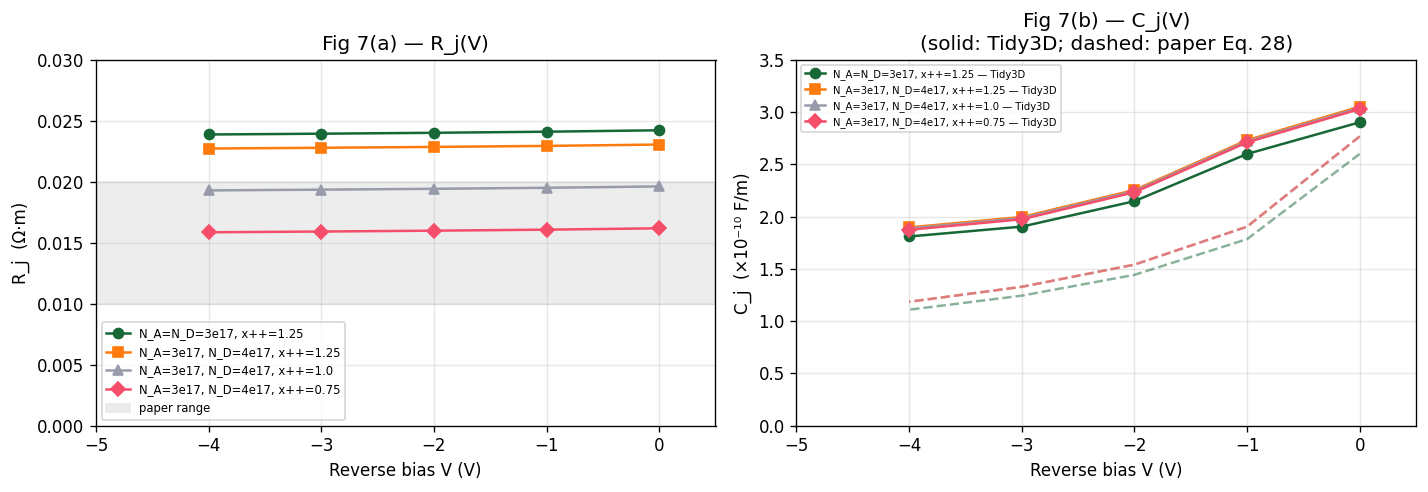}
\caption{Stage~2 carrier transport: \textbf{(a)} $R_j(V)$ and
\textbf{(b)} $C_j(V)$ for four doping splits matching Fig.~7 of
\citet{zhuang2024equivalent}. Dashed curves are the agent's port of the
paper's rib-only analytical formula (Eq.~28), within \textbf{4--11\%}
of the published values; solid curves are the full FDTD $C_j$ including
slab depletion at the metallurgical junction.}
\label{fig:mzm_charge}
\end{figure}

\paragraph{First attempt: agent set up the pipeline but missed the RF metrics.}
The agent built a single \PhotonForge\ parametric component
(Fig.~\ref{fig:mzm_layout}) that drove charge transport,
optical mode, and 3-D RF FDTD without re-export. Stage 2 (carrier transport,
Fig.~\ref{fig:mzm_charge}) matched the published analytical $C_j(V)$ from
Eq.~28 of \citet{zhuang2024equivalent} to within 4--11\% across the bias
sweep; the $\sim$30\% gap between the rib-only formula and the full FDTD
$C_j$ is real physical capacitance from slab depletion that the rib-only
formula does not capture. Stage 3 (RF FDTD on the loaded line) did not
match: at \SI{30}{\giga\hertz} the agent extracted
$Z_0 \approx \SI{15}{\ohm}$ against the paper's $\approx \SI{48}{\ohm}$,
$n_\mathrm{RF} \approx 3.1$ against $\approx 3.83$, and microwave loss
roughly twice the reference (Table~\ref{tab:mzm_rf}). The agent diagnosed
the two largest contributors in its own journal: (i) wave-port mode
mis-identification on the periodically loaded cross-section, where the
converged mode picked up a lossy higher-order CPS-substrate hybrid rather
than the intended quasi-TEM electrode mode, and (ii) a lumped-shunt
PN-junction admittance that double-counted the slab depletion already
captured by the FDTD.

\begin{figure}[!htbp]
\centering
\includegraphics[width=\textwidth]{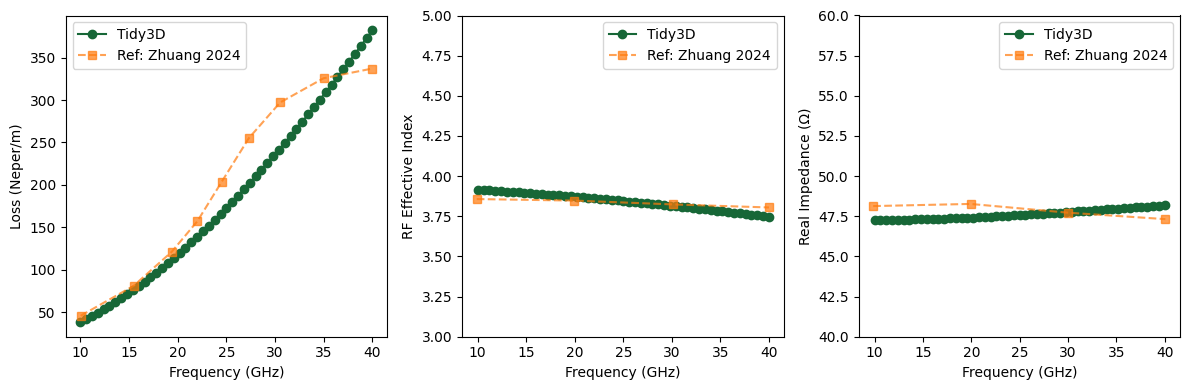}
\caption{Expert-reference loaded-line characteristics
\citep{khavasi2026cps}, overlaid on the Zhuang reference. Microwave loss
(left), RF effective index (centre), and characteristic impedance
(right) all land inside the acceptance windows of
Table~\ref{tab:mzm_rf}.}
\label{fig:mzm_amin_loaded}
\end{figure}

\begin{figure}[!htbp]
\centering
\includegraphics[width=0.85\textwidth]{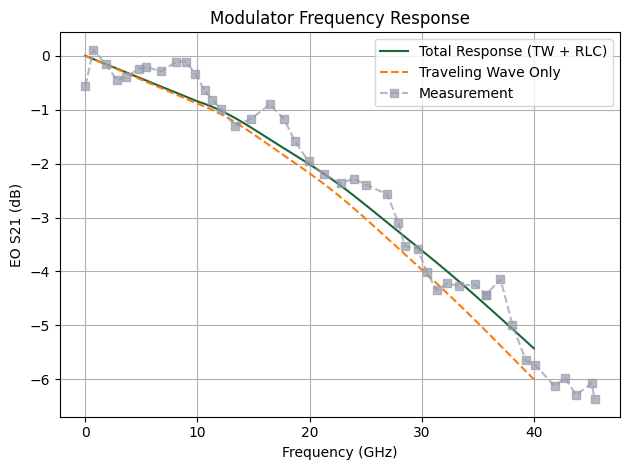}
\caption{End-to-end EO response from the expert reference. The
traveling-wave-only response (dashed) is multiplied by the lumped RLC
junction-divider transfer function to give the total response (solid),
which matches measured EO $S_{21}$ data from the McGill thesis of
Patel \citep{patel2015silicon} (squares) across
\SIrange{0}{45}{\giga\hertz}.}
\label{fig:mzm_amin_eo}
\end{figure}

\paragraph{Expert reference closes the RF stage.}
A co-author (A.K.) built a hand-coded reference of the same RF workflow on
the same Tidy3D solver stack \citep{khavasi2026cps}. It applied the two
fixes the autonomous run had diagnosed (long unloaded-CPS feeds with
ABCD-matrix de-embedding, and a distributed-shunt PN-junction admittance)
together with three additional setup choices the autonomous run had not
surfaced: mirror symmetry at the wave port, mesh refinement around the
loaded section, and per-arm junction $C_j$ and $R_j$ values calibrated to
the published loading (which the reference does not state numerically).
With those changes the loaded-line metrics all landed inside acceptance:
$Z_0 \in 47$--$48\,\Omega$, $n_\mathrm{RF} \in 3.85$--$3.92$, and
microwave loss within $\sim$10--15\% of the reference across band
(Fig.~\ref{fig:mzm_amin_loaded}). Composing the same pipeline with an
analytic traveling-wave EO transfer function and an RLC junction
voltage divider also reproduced the measured EO $S_{21}$ data from the
McGill thesis of Patel \citep{patel2015silicon} across
\SIrange{0}{45}{\giga\hertz} (Fig.~\ref{fig:mzm_amin_eo}), giving a
$\sim$\SI{26}{\giga\hertz} \SI{3}{\decibel} EO bandwidth.

\begin{figure}[!htbp]
\centering
\includegraphics[width=\textwidth]{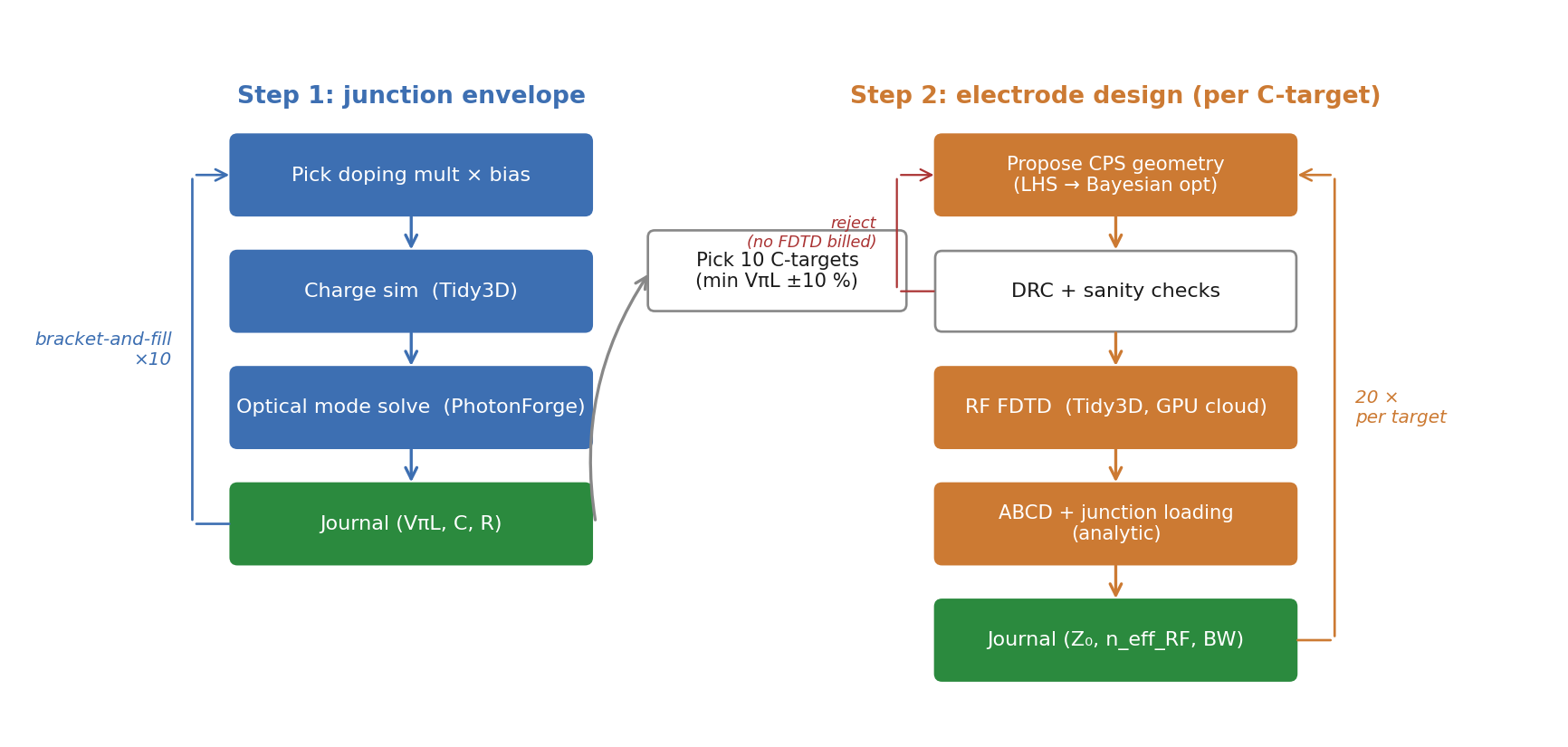}
\caption{Two-stage closed-loop architecture. Step~1 (blue) sweeps the
doping scaling factor and builds the junction $(V_\pi L, C)$ envelope;
Step~2 (orange) selects ten operating capacitances and runs an
independent 20-iteration Bayesian optimization of the eight-parameter
segmented-CPS electrode for each, gated by three DRC layers.}
\label{fig:amin_loop}
\end{figure}

\begin{figure}[!htbp]
\centering
\includegraphics[width=0.85\textwidth]{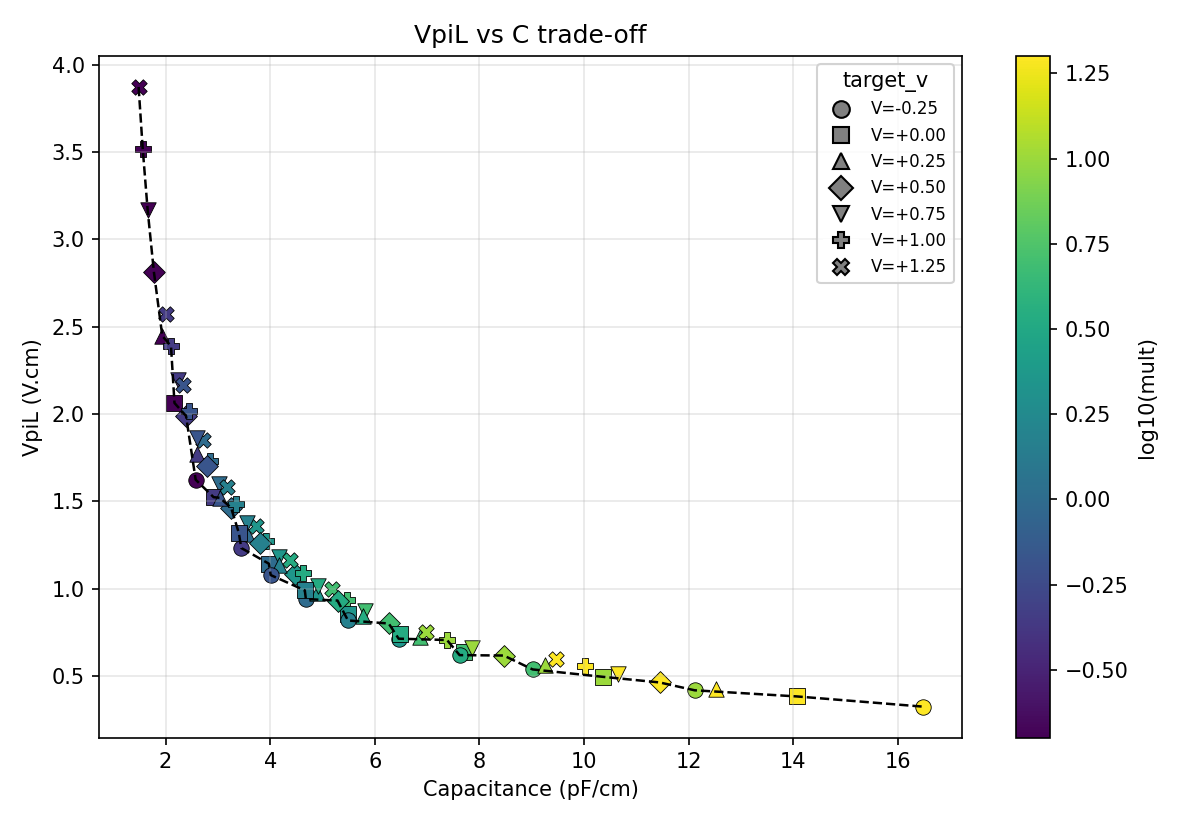}
\caption{Step-1 junction characterization on the $(V_\pi L, C)$ plane.
Colour encodes $\log_{10}\mathrm{mult}$; markers encode reverse bias.
The dashed line traces the lower envelope used by Step~2.}
\label{fig:amin_step1}
\end{figure}

\begin{figure}[!htbp]
\centering
\includegraphics[width=0.85\textwidth]{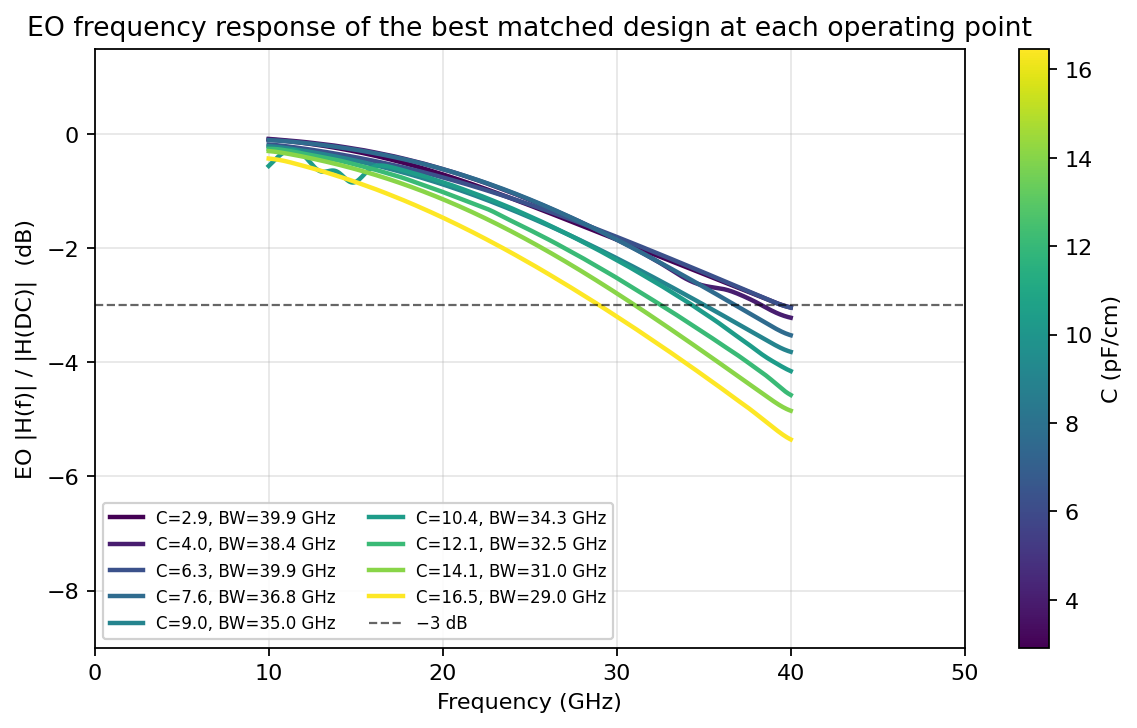}
\caption{EO $S_{21}$ magnitude of the best design at each of the ten
operating points, normalized to DC. Light-loading designs roll off near
\SIrange{39}{40}{\giga\hertz}; heavy-loading designs (under
$J_\mathrm{final}$) roll off near \SIrange{29}{37}{\giga\hertz} while
reaching $1/V_\pi L \gtrsim 3\,(\si{V}\cdot\si{cm})^{-1}$ on
sub-\SI{300}{\micro m} devices.}
\label{fig:amin_eo_s21}
\end{figure}

\paragraph{Closed-loop autonomous run on the corrected reference.}
Launched from that corrected reference notebook, a second closed-loop
agent ran automated multi-objective design optimization over the loaded
electrode under a hard 200-FDTD budget \citep{khavasi2026closedloop}. The
architecture was two stacked loops sharing one journal
(Fig.~\ref{fig:amin_loop}). Step~1 swept a doping scaling factor and
populated the lower envelope of the junction $(V_\pi L, C)$ plane
(Fig.~\ref{fig:amin_step1}); Step~2 selected ten operating capacitances
from that envelope and ran an independent 20-iteration Bayesian
optimization of the eight-parameter segmented-CPS electrode for each.
Three DRC layers (fab rules on the eight parameters, fixed process and
material stack, and setup-sanity checks including a minimum unloaded-CPS
pad length to keep the wave port off PML) gated every FDTD call. The run
produced nine distinct silicon MZM designs (ten operating points, two
sharing a Step-1 junction), all holding loaded $Z_0$ within $\pm 10\%$
of \SI{50}{\ohm} and tracing a bandwidth-efficiency frontier from
$\sim$\SI{40}{\giga\hertz} on a \SI{1.3}{mm} device to
$\sim$\SI{29}{\giga\hertz} on a \SI{282}{\micro m} device
(Table~\ref{tab:amin_frontier}); the corresponding EO $S_{21}$ family is
shown in Fig.~\ref{fig:amin_eo_s21}.

\begin{table}[!htbp]
\centering
\caption{Stage~3 headline metrics at \SI{30}{\giga\hertz}, agent
first-attempt vs.\ Zhuang reference. Reference values match the curve
in Fig.~\ref{fig:mzm_amin_loaded}.}
\label{tab:mzm_rf}
\begin{tabular}{lccc}
\toprule
Quantity (at \SI{30}{\giga\hertz}) & Reference paper & Agent first attempt & Acceptance window \\
\midrule
$Z_0$ (\si{\ohm})                       & 48   & 14.7--14.8 & $\pm 5\,\si{\ohm}$ \\
$n_\mathrm{RF}$                         & 3.83 & 3.10--3.13 & $\pm 2\%$ \\
$\alpha$ at \SI{30}{\giga\hertz} (Np/m) & 295  & 497--546   & $\pm 15\%$ \\
$\alpha$ at \SI{40}{\giga\hertz} (Np/m) & 340  & 607--646   & $\pm 15\%$ \\
\bottomrule
\end{tabular}
\end{table}

\begin{table}[!htbp]
\centering
\caption{The nine distinct silicon MZM designs from the closed-loop
run, sorted by $V_\pi L$. $L_\mathrm{MZM}$ is the length at which
\SI{5}{\decibel} extinction at \SI{2}{\volt} push--pull is met;
$Z_{0,\mathrm{loaded}}$ and $n_\mathrm{RF}$ are reported at the
Step-2 cost-function frequency; BW is the EO \SI{3}{\decibel}
bandwidth.}
\label{tab:amin_frontier}
\small
\begin{tabular}{rrrrrrr}
\toprule
$C$ [pF/cm] & $V_\pi L$ [V$\cdot$cm] & $1/V_\pi L$ [(V$\cdot$cm)$^{-1}$] & $L_\mathrm{MZM}$ [$\mu$m] & $Z_{0,\mathrm{loaded}}$ [$\Omega$] & $n_\mathrm{RF}$ & BW$_{3\mathrm{dB}}$ [GHz] \\
\midrule
2.92  & 1.523 & 0.66 & 1325 & 52.2 &  3.79 & 39.9 \\
4.01  & 1.078 & 0.93 &  937 & 54.1 &  4.40 & 38.4 \\
6.27  & 0.800 & 1.25 &  696 & 49.2 &  5.75 & 39.9 \\
7.62  & 0.619 & 1.62 &  538 & 54.9 &  6.67 & 36.8 \\
9.02  & 0.537 & 1.86 &  467 & 51.0 &  7.16 & 35.1 \\
10.35 & 0.495 & 2.02 &  430 & 52.2 &  8.24 & 34.3 \\
12.11 & 0.418 & 2.39 &  363 & 53.0 &  9.40 & 32.5 \\
14.07 & 0.383 & 2.61 &  333 & 52.6 & 10.71 & 31.0 \\
16.47 & 0.324 & 3.09 &  282 & 49.1 & 11.35 & 29.0 \\
\bottomrule
\end{tabular}
\end{table}

\bibliographystyle{unsrtnat}
\bibliography{references}

\end{document}